\documentclass[fleqn,usenatbib,twocolumn]{mnras}

\usepackage[T1]{fontenc}
\usepackage{ae,aecompl}

\usepackage{etoolbox}
\makeatletter
\patchcmd\@combinedblfloats{\box\@outputbox}{\unvbox\@outputbox}{}{%
}%
\makeatother

\usepackage{graphicx}	
\usepackage{amsmath}	
\usepackage{amssymb}	
\usepackage[usenames, dvipsnames]{color} 
\usepackage{float}
\usepackage{verbatim}
\usepackage{orcidlink}
\usepackage{xcolor}
\usepackage{newtxtext,newtxmath}

\newcommand{\civ}{\ion{C}{iv}}
\newcommand{\civline}{\ion{C}{iv}~$\lambda$1550}
\newcommand{\sivline}{\ion{S}{iv}~$\lambda$1400}

\newcommand{\mgiiline}{\ion{Mg}{ii}~$\lambda$2800}
\newcommand{\heiiuv}{\ion{He}{ii}~$\lambda$1640}
\newcommand{\ciiiline}{\ion{C}{iii}]~$\lambda$1909}
\newcommand{\heii}{\ion{He}{ii}}
\newcommand{\nv}{\ion{N}{V}}
\newcommand{\nvline}{\ion{N}{V}~$\lambda$1240}
\newcommand{\oiline}{\ion{O}{i}~$\lambda$6300}
\newcommand{\oiiiline}{\ion{O}{iii}~$\lambda\lambda$4960,5008}
\newcommand{\view}{15} 

\title
[A disc wind model for blueshifts in quasar lines]
{
A disc wind model for blueshifts in quasar broad emission lines
}

\author[J. H. Matthews et al.]
{James~H.~Matthews$^{\orcidlink{0000-0002-3493-7737}}$,$^{1}$\thanks{james.matthews@physics.ox.ac.uk}
Jago Strong-Wright$^{\orcidlink{0000-0002-7174-5283}}$,$^{2,3}$
Christian Knigge$^{\orcidlink{0000-0002-1116-2553}}$,$^{4}$
Paul Hewett$^{\orcidlink{0000-0002-6528-1937}}$,$^{3}$
\newauthor
Matthew J. Temple$^{\orcidlink{0000-0001-8433-550X}}$,$^{5}$ 
Knox S. Long$^{\orcidlink{0000-0002-4134-864X}}$,$^{6,7}$ 
Amy L. Rankine$^{\orcidlink{0000-0002-2091-1966}}$,$^{8}$
Matthew Stepney$^{\orcidlink{0000-0002-7711-0537}}$,$^{4}$
Manda Banerji$^{\orcidlink{0000-0002-0639-5141}}$$^{4}$ \newauthor
and
Gordon T. Richards$^{\orcidlink{0000-0002-1061-1804}}$$^{9}$
\\$^1$Department of Physics, Astrophysics, University of Oxford, Denys Wilkinson Building, Keble Road, Oxford, OX1 3RH, UK
\\$^2$DAMTP, Centre for Mathematical Sciences, University of Cambridge, Wilberforce Road, Cambridge CB3 OWA, UK
\\$^3$Institute of Astronomy, University of Cambridge, Madingley Road, Cambridge, CB3 0HA, UK
\\$^{4}$School of Physics \& Astronomy, University of Southampton, Southampton SO17 1BJ, UK
\\$^{5}$Instituto de Estudios Astrof\'{\i}sicos, Universidad Diego Portales, Av. Ej\'ercito Libertador 441, Santiago 8370191, Chile\\
$^{6}$Space Telescope Science Institute, 3700 San Martin Drive, Baltimore, MD, 21218, USA\\
$^{7}$Eureka Scientific Inc., 2542 Delmar Avenue, Suite 100, Oakland, CA, 94602-3017, USA\\
$^{8}$Institute for Astronomy, University of Edinburgh, Royal Observatory, Blackford Hill, Edinburgh EH9 3HJ, UK\\
$^{9}$Department of Physics, Drexel University, 32 S. 32nd Street, Philadelphia, PA 19104, USA\\
}

\date{Accepted 2023 September 18. Received 2023 August 14}

\pubyear{2023}

\begin{document}
\label{firstpage}
\pagerange{\pageref{firstpage}--\pageref{lastpage}}
\maketitle

\begin{abstract}
Blueshifts – or, more accurately, blue asymmetries – in broad emission lines such as C\textsc{iv} $\lambda$1550 are common in luminous quasars and correlate with fundamental properties such as Eddington ratio and broad absorption line (BAL) characteristics. However, the formation of these blueshifts is still not understood, and neither is their physical connection to the BAL phenomenon or accretion disc. In this work, we present Monte Carlo radiative transfer and photoionization simulations using parametrized biconical disc-wind models. We take advantage of the azimuthal symmetry of a quasar and show that we can reproduce C\textsc{iv} blueshifts provided that (i) the disc-midplane is optically thick out to radii beyond the line formation region, so that the receding wind bicone is obscured; and (ii) the system is viewed from relatively low (that is, more face-on) inclinations ($\lesssim40^\circ$). We show that C\textsc{iv} emission line blueshifts and BALs can form in the same wind structure. The velocity profile of the wind has a significant impact on the location of the line formation region and the resulting line profile, suggesting that the shape of the emission lines can be used as a probe of wind-driving physics. While we are successful at producing blueshifts/blue asymmetries in outflows, we struggle to match the detailed shape or skew of the observed emission line profiles. In addition, our models produce redshifted emission-line asymmetries for certain viewing angles. We discuss our work in the context of the C\textsc{iv} $\lambda$1550 emission blueshift versus equivalent-width space and explore the implications for quasar disc wind physics. 
\end{abstract}

\defcitealias{richards_unification_2011}{R11}
\defcitealias{rankine_bal_2020}{R20}
\defcitealias{matthews_stratified_2020}{M20}
\newcommand{\python}{\textsc{Python}}

\section{Introduction}
Luminous active galactic nuclei (AGN) and quasars produce outflows, in the form of collimated, relativistic jets and mass-loaded winds. Both of these classes of outflows are intimately connected to the accretion disc and have profound implications for our understanding of black holes, particularly their growth and evolution. In addition, outflows are invoked as possible feedback agents \citep[see][for reviews]{fabian_observational_2012,morganti_many_2017,harrison_agn_2018}, providing a mechanism through which the vast accretion energy released close to the black hole can couple to the surroundings. 

AGN winds -- our focus here -- are capable of expelling gas and/or quenching star formation in their host galaxies \citep[e.g.][]{silk_quasars_1998,king_black_2003,di_matteo_energy_2005,feruglio_quasar_2010,hopkins_quasar_2010,costa_feedback_2014}, and are critical ingredients in cosmological simulations \citep[e.g.][]{beckmann_cosmic_2017,chisari_impact_2018}. In addition, the impact of these winds can feasibly account for observed scaling relations such as the $M$--$\sigma$ relation \citep{silk_quasars_1998,king_black_2003,di_matteo_energy_2005}. Despite the clear importance and apparent ubiquity of outflowing material, the driving mechanism of the winds and the precise relationship between the outflows and the accretion process both remain poorly understood. 

Approximately $20 - 40$ per cent  of quasars show blue-shifted broad absorption lines (BALs) in their rest-frame ultraviolet (UV) spectra \citep{weymann_comparisons_1991,hewett_frequency_2003,knigge_intrinsic_2008,allen_strong_2011,dai_intrinsic_2012}. These BAL quasars provide unambiguous evidence of an absorbing outflow along the line-of-sight. The general paradigm is that BALs are produced by an accretion disc wind that intersects the observer's sightline. A fairly common geometry is an equatorial wind \citep[e.g.][]{murray_accretion_1995}, roughly consistent with simulations of line-driven winds \citep{proga_dynamics_2004}. However, this equatorial disc-wind geometry is hard to reconcile with the apparent similarity in spectral properties between BAL and non-BAL quasars \citep[e.g.][]{weymann_comparisons_1991,matthews_quasar_2017,yong_using_2018,rankine_bal_2020}. 
Furthermore, an equatorial viewing angle would lead BALs to be under-represented in quasar samples due to suppression of the continuum flux by an obscuring torus, foreshortening of the accretion disc and/or limb darkening \citep{krolik_what_1998,matthews_quasar_2017}. In addition to questions of geometry, measuring distances to BAL formation regions throws up further puzzles -- how can apparent distances of $\sim$kpc scales (in many cases) inferred from density sensitive lines \citep[e.g.][]{borguet_major_2012,arav_evidence_2018,xu_vltx-shooter_2018,leighly_physical_2022} be reconciled with the much smaller launching radii expected from line-driven disc wind simulations \citep[e.g.][]{proga_dynamics_2004}? The former inferred BAL distances are much larger than the scale of the quasar accretion disc, which has a self-gravitation radius of $\sim 0.1~{\rm pc}$. All in all, although BALs must be formed in quasar outflows of some form, whether these outflows are actual `disc winds` in the conventional sense is not at all clear. Indeed, perhaps BAL quasars are not a single, monolothic class of objects, and the physics driving their observational appearance could be as heterogeneous as the spectral signatures themselves. 

Although blue-shifted BALs are the `smoking gun' signature of an absorbing outflow, a disc wind can also affect the spectrum in other ways. 
For example, various authors have proposed a common origin for the formation of broad emission lines and BALs within a biconical wind structure \citep[e.g.][]{emmering_magnetic_1992,de_kool_radiation_1995,murray_accretion_1995,elvis_structure_2000}. Emission line asymmetries can also be produced in outflows, but these are more challenging to interpret than BALs. While we know that the emission from the blue wing comes from material moving towards us, this is subtly different from knowing it is formed in an outflow.  Assuming symmetry about the disc axis, an outflow also has a receding emitting volume. Whereas in a pure scattering atmosphere with spherical symmetry a classical P-Cygni profile is produced, for the more complex quasar system the observed profile depends on a number of factors; these factors include the covering factor of the flow, the emissivity profile of the line, the opacity of the midplane and the interplay between inclination and wind launching angle. Indeed, the early papers describing the discovery of \civline\ emission line blueshifts discussed formation in a quasar outflow \citep{gaskell_redshift_1982,wilkes1984}, but alternative models in which the BLR is inflowing have also been proposed \citep{gaskell_line_2013,gaskell_case_2016}.

Extending the `Eigenvector I' paradigm \citep{boroson_emission-line_1992,sulentic_eigenvector_2000,sulentic_average_2002,marziani_searching_2001,marziani_searching_2003} to higher redshift, \citet{sulentic_c_2007} and \citet[][hereafter \citetalias{richards_unification_2011}]{richards_unification_2011} identified the fundamental nature of the \civ\ equivalent width (EW) versus blueshift parameter space (hereafter {\em \civ\ emission space}). \citetalias{richards_unification_2011} demonstrated the existence of fundamental trends across this parameter space in terms of emission line EWs, Eddington ratio and radio loudness. \citet[hereafter \citetalias{rankine_bal_2020}]{rankine_bal_2020} expanded on this by using mean-field independent component analysis to reconstruct BAL and non-BAL quasar spectra, showing that the BAL quasars appear to be drawn from the same parent population as non-BAL quasars. The results presented by \citetalias{richards_unification_2011} and  \citetalias{rankine_bal_2020} have important implications for our understanding of quasar outflows; they show that the strongest, broadest BAL troughs are formed in quasars with high \civ\ blueshifts. The results can be interpreted within the framework of line-driven winds \citep{proga_dynamics_2000,proga_dynamics_2004,proga_theory_2005}, with high blueshift quasars corresponding to high Eddington fraction quasars with soft SEDs that provide the optimal conditions for line-driving \citep{rivera_exploring_2022,temple_testing_2023}. However, it is important to remember that \civ\ blueshift versus EW space is likely a projection of a multi-dimensional space, and there are many questions that arise -- for example, why are BALs only seen in most, but not all, of this parameter space? What is the role of inclination? How are the \civ\ blueshifts formed in the first place?  

Radiative transfer calculations with varying degrees of complexity have been successful in simulating emission lines at observable levels from quasar disc wind models \citep{murray_accretion_1995,waters_reverberation_2016,matthews_testing_2016,matthews_stratified_2020,naddaf2022}. A few authors have also specifically discussed synthetic emission line asymmetries and blueshifts in more detail. For example, \cite{chajet_magnetohydrodynamic_2013,chajet_magnetohydrodynamic_2017} used the magneto-hydrodynamic (MHD) wind model of \cite{emmering_magnetic_1992}, a slightly modified version of the classic \cite{blandford_hydromagnetic_1982} self-similar solution, to study line widths and shapes as a function of the wind geometry. \cite{yong_kinematics_2017} also studied line asymmetries using the same kinematic prescription from \cite{shlosman_winds_1993} adopted here, including a simplified treatment of radiative transfer. They found they could produce blueshifts in their models, and that higher blueshifts were produced for polar winds and polar viewing angles due to velocity projection effects and the obscuring effect of the disc midplane. Although the observed blue asymmetries are most common in the \civline\ line, they can be seen in other emission lines. For example, \cite{gravity_collaboration_spatially_2020} reported an asymmetric Paschen-$\alpha$ line in IRAS~09149-6206, also noting the importance of an opaque midplane in their modelling.  

An interesting point of astrophysical comparison is protoplanetary systems, in which relatively low temperature discs of gas and dust form around low-mass stars early in their lives \citep{williams_protoplanetary_2011}. In these objects, blueshifted emission lines are observed in forbidden lines such as [Ne \textsc{ii}]\,12.81\,$\mu$m and \oiline. These blueshifted profiles are attributed to line formation in a disc wind, which, in T Tauri systems, is thought to be photoevaporative or thermally driven \citep{pascucci_evolution_2020}, although there may also be an inner MHD wind. The temperatures and kinematics of these winds are rather different from the putative broad-line region (BLR) or BAL winds, with typical velocities of $\sim 10~{\rm km~s}^{-1}$ and electron temperatures of $\sim3000$\,K. \citep{banzatti_kinematic_2019}. The thermal winds are produced when the sound speed exceeds the local escape speed, and -- though the heating and cooling physics is different -- the winds bear similarity to the Compton-heated thermal winds thought to occur in X-ray binaries \citep{begelman_compton_1983,done_thermal_2018,higginbottom_radiation-hydrodynamic_2018} and possibly AGN \citep{mizumoto_thermally_2019,ganguly_synthetic_2021}. A number of simulations of thermal/photoevaporative winds in protoplanetary discs have been successful in producing blueshifted profiles comparable to those observed \citep[e.g.][]{sellek_general_2021}. In these simulations, the midplane of dust and cold gas is responsible for obscuring the receding wind bicone; this scenario is very similar to the one we will discuss for blueshift formation in quasar winds. 

\begin{figure*}
    \centering
    \includegraphics[width=1.0\linewidth]{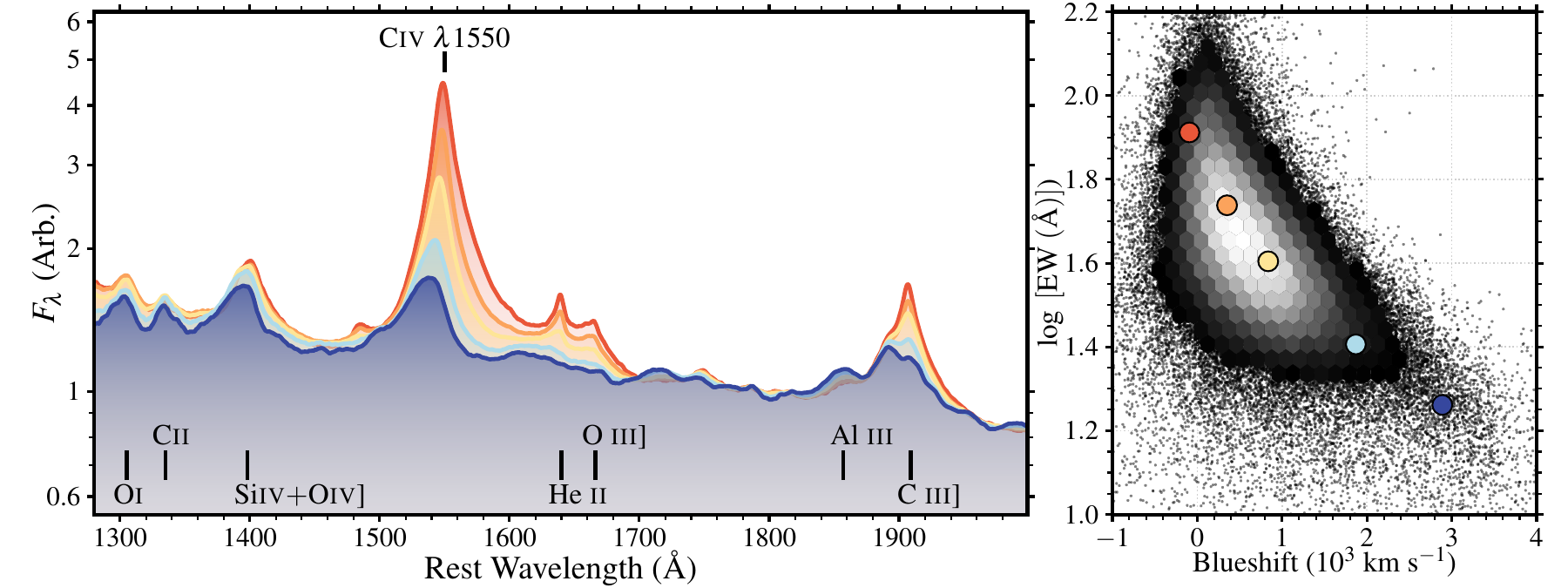}
    \caption{A graphical summary of quasar spectra as a function of \civ\ emission line properties.
    {\sl Left:} Composite spectra of quasars at five locations in the \civ\ blueshift and equivalent-width space over the wavelength range $1280-2000$\AA, so as to show both the \civline\ resonance line and the \ciiiline\ emission complex. The composite spectra are normalised to the mean flux in the 1760\AA\ to 1840\AA\ wavelength range. A high EW in \civ\ is accompanied by symmetric \civ\ lines and strong \heii\ emission, whereas systems with blueshifted \civ\ have lower \civ\ EWs and weaker \heii. As shown by \citetalias{richards_unification_2011} and \citetalias{rankine_bal_2020}, this trend occurs systematically across \civ\ emission space. {\sl Right:} locations in \civ\ emission space of the five composite spectra compared to the background distribution of quasars with $z<2.8$ from \citetalias{rankine_bal_2020}.}
 \label{fig:composite}
\end{figure*}

In this paper, we expand on previous efforts to model quasar emission lines (and their asymmetries) by conducting Monte Carlo radiative transfer and photoionization simulations in an azimuthally symmetric biconical disc wind geometry. Our approach allows us to include a more physical and self-consistent treatment of the line formation mechanism compared to previous studies. We are also able to calculate the emissivity of the wind plasma, which can be used to infer information about the power and mass loss rates of a disc wind associated with a \civ\ blueshift and study where the emission lines are actually forming in the wind. Our aim is not to model a population of quasars, or to reproduce the full diversity of quasar spectra. Instead, we will investigate whether \civ\ blueshifts can be produced for reasonable model assumptions within this assumed geometry, and whether they can explain the `windy' component that is apparently needed to explain the diversity of quasar spectra. Our paper is structured as follows. We first introduce the observational properties of the \civ\ emission line and our numerical method in Section~\ref{sec:method}. Following an heuristic demonstration of our approach in Section~\ref{sec:heuristic} we present a series of simulated synthetic spectra in Section~\ref{sec:results} and explore trends with the physical parameters and geometry of the wind model. In Section~\ref{sec:discuss}, we discuss our results in the context of quasar feedback, disc wind theory and line asymmetries in a variety of astrophysical settings, before concluding in Section~\ref{sec:conclusions}. Throughout, we follow convention by using the shorthand \civline\ to refer to the \civ\ doublet (corresponding to the resonant 1s$^2$2p$\to$1s$^2$2s transitions) with rest frame vacuum wavelengths $1548.19$ and $1550.77$\AA\ \citep[according to the NIST Atomic Spectra Database;][]{nist}.


\section{Observational Data and Numerical Methods}
\label{sec:method}
\subsection{Observed Spectra}
\label{sec:observed}
Before introducing our numerical methods and discussing the physics of line formation in disc winds, we first review what is known observationally about emission line asymmetries. As stated above, \civ\ blueshifts were discovered some four decades ago \citep{gaskell_redshift_1982}. We now know that, for typical quasars at $z\sim 1.5-3.5$, the \civline\ emission line profiles form a continuum from  strong and symmetric to weak, highly asymmetric and blueshifted. 

The results of the simulations in the paper are presented primarily as \civline\ emission-line profiles. As motivation and orientation for the investigation we use the spectroscopic sample from the Sloan Digital Sky Survey (SDSS) employed in \citetalias{rankine_bal_2020}. In particular, we will use this sample as the basis for presenting composite spectra and plotting the distribution of \civline\ EWs and blueshifts. 

To quantify the asymmetry in the emission line profiles, we follow the recipe of \citet{coatman_2016} to measure the EW and define blueshift of a line as
\begin{equation}
    {\rm blueshift} = -c\frac{\lambda_{\rm half} - \lambda_0}{\lambda_0} .
\label{eq:blueshift}
\end{equation}
Here, $\lambda_{\rm half}$ is the wavelength that bisects the integrated line flux, and $\lambda_0$ is the line centre wavelength (1549.5\AA\ for the \civ\ UV resonance line doublet). This definition is positive if the line is blueshifted, opposite to the usual radial velocity definition. 
We also define the `skew' of the line profile as the relative offset between the velocity of the emission line peak and the blueshift, i.e.
\begin{equation}
    {\rm skew} = -\left(c\frac{\lambda_{\rm peak} - \lambda_0}{\lambda_0}\right) - {\rm blueshift},
\label{eq:skew}
\end{equation}
where $\lambda_{\rm peak}$ is the wavelength of the peak line flux. We will use these two quantities to quantify of the asymmetry in our simulations and compare to the observed data, while using the skew as a measure of the line shape. This definition of skewness leads to a negative value when the line peak is redward of $\lambda_{\rm half}$.

Fig.~\ref{fig:composite} (right-panel) shows the \civ\ emission space -- that is, the observed distribution of blueshift and EW for the \civ\ emission-line -- in the quasar sample from \citetalias{rankine_bal_2020}. For five locations in this parameter space, chosen so as to span a range of \civ\ emission-line properties, the left-hand panel shows high signal-to-noise ratio composite spectra (each generated using a minimum of 500 quasar spectra). The full set of 33 composite spectra, from which these five are drawn, are described by \cite{stepney}. The significant range of line profiles for \civline\ and the \ciiiline\ emission-line complex\footnote{The emission present at $\sim 1830-1920$\,\AA \ is a blend of {\ion{Al}{iii}}$ \lambda\lambda$1855,1863, {\ion{Si}{iii}}] $\lambda$1892, \ion{C}{iii}] $\lambda$1909 and the \ion{Fe}{iii} UV 34 multiplet.} is striking. The trend originally reported by \citetalias{richards_unification_2011} can be clearly seen: the high-blueshift quasars have weak \heiiuv\ emission and weaker \ciiiline\ compared to the low-blueshift `peaky' spectra. Additionally, the highly blueshifted, lowest EW \civ\ line in the composite spectra in Fig.~\ref{fig:composite} has a negative skew. In fact, negative skews when blueshifts are present in the \civline\ are a general feature of SDSS quasar spectra (e.g. \citetalias{richards_unification_2011}; \citetalias{rankine_bal_2020}; \citealt{stepney}; see also Section~\ref{sec:skew}). 

\subsection{Monte Carlo Radiative Transfer and Photoionization}
We use a well-tested Monte Carlo radiative transfer (MCRT) and photoionization code, confusingly called \python, to simulate rest-frame UV spectra from a biconical disc wind geometry. The code was originally introduced by \cite{long_modeling_2002}, but has since been updated and upgraded  \citep{sim_modelling_2005,higginbottom_simple_2013,matthews_testing_2016,matthews_disc_2016}. Radiative transfer is treated using the Monte Carlo method, with line transfer treated in the Sobolev limit \citep[see][for a review]{noebauer_monte_2019}. During a simulation, the code follows energy packets of discretised radiation (hereafter `$r$-packets', following the notation of \citealt{lucy_monte_2002}) as they make their way through a user-defined geometry or imported grid. The frequency distribution of these packets and their overall energy normalisation are set by the input SED (Section~\ref{sec:photons}). The code first runs a series of `ionization cycles', during which Monte Carlo estimators are used to iteratively calculate the temperature, ionization state and emissivity of the wind. Once a satisfactorily converged solution is obtained (see section~\ref{sec:results}), the temperature and ionization structure is held fixed. The synthetic spectrum is then created over a limited wavelength range (`spectral cycles') for a number of specified observer viewing angles. 

\subsubsection{Recent code improvements}
Our MCRT code has been improved in a number of ways since it was last used to simulate quasar spectra. These improvements make only small quantitative differences to the spectrum observed from a quasar model. The first development since our work in \citet[][hereafter \citetalias{matthews_stratified_2020}]{matthews_stratified_2020} is that we have improved our treatment of frame transformations and Doppler shifts so as to fully account for special relativistic effects. In particular, we now explicitly conserve energy in the co-moving frame and include the Lorentz factor in, for example, the calculation of the number density in a given volume. We also include all relevant special relativistic effects (such as relativistic aberration) when calculating photon frequencies. 

An additional improvement has been made to the macro-atom mode of operation, in which transition probabilities are sampled to determine how energy flows into and out of the excitation, ionization and kinetic energy pools. Previously, this was achieved by Monte Carlo sampling of the transition probabilities as originally proposed by \cite{lucy_monte_2002,lucy_monte_2003}. Now, we are able to dramatically speed up this section of the code by using standard Markov chain techniques \citep{ross}, as applied to the MCRT macro-atom scheme by \cite{ergon_monte-carlo_2018}.

\subsubsection{Wind model and clumping}
As in our previous works, we use a flexible parametrized model for a biconical wind, originally introduced by \cite{shlosman_winds_1993} as a model for disc winds in accreting white dwarf system. In this prescription, the wind rises from an accretion disc with straight poloidal streamlines between minimum and maximum radii, $r_{\rm min}$ and $r_{\rm max}$, with a specified velocity law and range of opening angles. 
The velocity at a given distance $l$ along a streamline is given by
\begin{equation}
    v(l) = v_0 + (v_\infty - v_0) \left[ \frac{(l/R_v)^{\alpha}}{1+(l/R_v)^{\alpha}} \right],
    \label{eq:velocity}
\end{equation}
and is controlled by four parameters: the acceleration length, $R_v$, the acceleration exponent, $\alpha$, and the initial and terminal velocities, $v_0$ and $v_\infty$. The rotational velocity, $v_\phi$, is initially set to the Keplerian velocity at the streamline base, and specific angular momentum is then conserved along streamlines. The angles that the poloidal streamlines make with the vertical $z$-axis are spaced evenly between the inner and outer opening angles of the wind, $\theta_{\rm min}$ and $\theta_{\rm max}$, respectively. A mass-loss rate, $\dot{M}_w$, is imposed and the local density follows from mass conservation. The model is azimuthally symmetric around the polar axis, and has reflective symmetry around the disc midplane. We conduct our simulations on a cylindrical ($x,z$) grid with the wind discretised into logarithmically spaced cells. The number of cells is chosen to adequately resolve the dense wind base, and ensure cells do not have excessively large velocity changes or continuum optical depth. 

As in our previous work, we use a simple treatment of clumping, inspired by modelling of stellar winds, known as `microclumping' \citep[e.g.][]{hamann_spectral_1995,matthews_testing_2016}. Clumping is needed in order to prevent over-ionization of the flow; without it, emission lines are only formed if either the X-ray source is weaker than expected for a quasar of this luminosity \citep{higginbottom_simple_2013,matthews_testing_2016}, or the mass-loss rate through the wind is significantly higher than the accretion rate through the disc. In fact, we have to make a few assumptions in order to get the wind to emit strongly at all, which we discuss further in Sections~\ref{sec:ew} and \ref{sec:limitations}. 

\begin{figure*}
    \centering
    \includegraphics[width=1.0\linewidth]{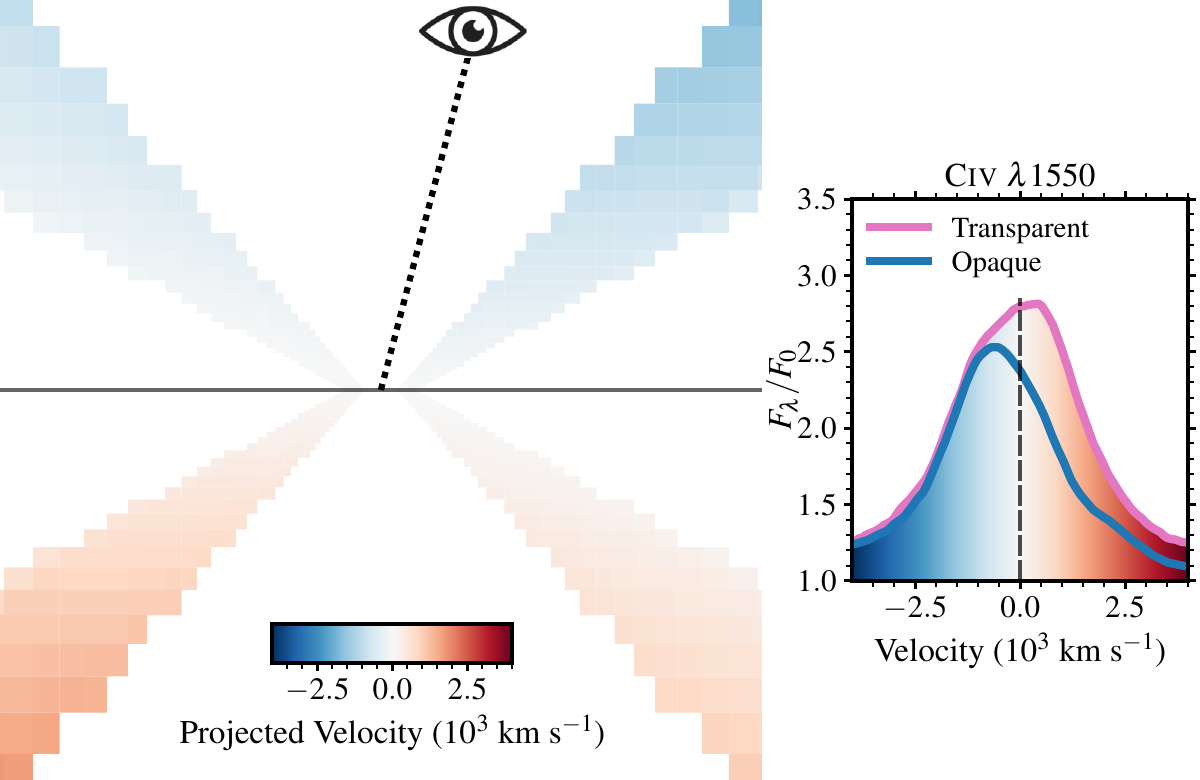}
    \caption{A heuristic demonstration of blueshift formation in an axisymmetric, biconical wind. The left-hand panel shows the wind geometry for the fiducial model described in the text (model n), and the colour denotes the projected outflow velocity for an observer at $\theta_i=\view^\circ$. The right-hand panel shows the observed (synthetic) \civline\ emission line profile from a MCRT photoionization calculation using the same geometry, with and without an opaque midplane. With an opaque midplane, the receding wind cone is obscured and a clear blueshifted asymmetry is produced.}
    \label{fig:heuristic}
\end{figure*}

\subsubsection{Radiation sources and sinks}
\label{sec:photons}
The $r$-packets in our MCRT modelling are all launched from a central, isotropic source, with a frequency distribution that matches the SED used by \citetalias{matthews_stratified_2020}. This SED shape is a combination of a multi-temperature, blackbody accretion disc with a standard \cite{shakura_reprint_1973} temperature profile, and a high energy power-law with spectral index $0.9$ (that is, $F_\nu \propto \nu^{-0.9}$). We explore the sensitivity of our results to the adopted SED and its anisotropy in Section~\ref{sec:sed_impact}, including a simulation in which the disc component of the radiation field is anisotropic and samples the emissivity profile of a multi-temperature accretion disc. The $r$-packets are propagated through the simulation grid until they escape the system or strike an opaque surface; if the latter occurs, the $r$-packet is discarded and any heating effect is not accounted for in this work. 
The inner $6~r_g$ (where $r_g$ is the gravitational radius), is always a purely absorbing opaque surface in our simulations. This choice corresponds to the innermost stable circular orbit of a non-spinning black hole. The midplane is either set to be transparent or opaque, with a parameter $R_{\rm mid}$ that sets the horizontal distance at which this midplane becomes transparent again. An opaque midplane is crucial for forming the blueshifts we observe. 

\section{Heuristic Demonstration}
\label{sec:heuristic}
We begin by using a single MCRT simulation of a biconical disc wind to demonstrate the basic principle underlying the blueshift formation in our models. We use a model comparable to Model A from \citetalias{matthews_stratified_2020}, with full model parameters given in Table~\ref{grid_table}. Broadly speaking, the model is designed to mimic a typical quasar in terms of its mass and luminosity. The wind has a mass-loss rate equal to the accretion rate and is clumpy, with a filling factor of $f_V = 0.01$. The wind is launched between $300$ and $600~r_g$, and has an inner wind opening angle of $\theta_{\rm min}=45^\circ$, a covering factor of $\Omega/4\pi=0.2$, and an acceleration exponent of $\alpha=1$. This model is denoted model n) in subsequent discussions and figures.

Fig.~\ref{fig:heuristic} shows a representation of how blueshifted asymmetries can form in this axisymmetric system. The figure is intended as heuristic or illustrative, but it is not a schematic diagram in the sense that the plotted colourmaps and synthetic spectra are obtained from the full, self-consistent MCRT photoionization calculation. The wind geometry is shown, colour-coded by projected velocity for an observer at $\theta_i=\view^\circ$. We discuss the inclination dependence further in Section~\ref{sec:inclination}. The \civline\ line profile is also plotted, and shows the results with and without an obscuring midplane disc (OMD). A clear blue-shifted asymmetry can be seen in the opaque midplane case, as the receding wind cone is obscured. There is a slight redshifted asymmetry in the case with a transparent midplane due to slight preferentially absorption of the blue wing by the wind itself. 

Fig.~\ref{fig:heuristic} provides a heuristic, but quantitative, demonstration of the basic principle of obscuring the receding part of the wind to break the symmetry of the system (and of the line profile). This general mechanism for forming line profile asymmetries -- obscuration of the receding part of a quasar outflow, possibly by an extended disc-like structure -- was discussed even in the earliest studies of \civ\ blueshifts \citep{gaskell_redshift_1982,wilkes1984}. Since then, this scenario has been explored in more detail using radiative transfer, albeit with simpler methods than employed here \citep{chajet_magnetohydrodynamic_2013,chajet_magnetohydrodynamic_2017,yong_kinematics_2017,gravity_collaboration_spatially_2020}, and similar proposals have been made for the line asymmetries in protoplanetary discs \citep{sellek_general_2021}. A related model was discussed by \cite{richards_broad_2002}, in which the midplane was only opaque at high inclinations due to limb opacity / path length effects. Similarly, blueshifts in AGN narrow lines such as \oiiiline\ have been attributed to, and modelled with, biconical winds on larger ($\sim 100{\rm s}$ of pc) narrow-line region scales with circumnuclear or midplane dust obscuring the receding bicone \citep{crenshaw_radial_2010,bae_prevalence_2016}.

The visual representation in Fig.~\ref{fig:heuristic} is necessarily simplified, because although the wind structure has azimuthal symmetry the observer also sees imprints on the line profile from the rotational velocity. The relative importance of rotation in determining the line profile shape depends on the location of the line formation region in velocity space (see e.g. Section~\ref{sec:velocity_bals}). The effect of the rotational velocity on the line profile is discussed explicitly in Section~\ref{sec:impacts}.

\begin{table}
\centering
\begin{tabular}{p{1.5cm}p{3cm}p{1.2cm}p{1.2cm}}
\hline 
Parameter & Grid Points & Model e) & Model n)  \\
\hline \hline 
$\dot{M}_w (M_{\odot}~\mathrm{yr}^{-1})$ 	& ($1,5,10$) & $5$ & $5$\\
$\theta_{\rm min}$ 	& ($20^{\circ},45^{\circ},70^{\circ}$) & $45^{\circ}$ & $45^{\circ}$ \\ 
$\alpha$ & ($0.5,1,1.5$) & $0.5$ & $1$ \\
$R_v~(\rm cm)$ 	& ($10^{17},10^{18},10^{19}$) & $10^{19}$ & $10^{19}$  \\ 
$f_\infty$ & ($1,2,3$) & $1$ & $1$ \\
\hline
\multicolumn{4}{|l|}{Important Fixed or Derived Parameters} \\ \hline
Parameter & Value & \multicolumn{2}{|l|}{Description}  \\
\hline \hline
$M_{\mathrm{BH}}$ 	   &	$10^9~{\rm M_\odot}$                                  & \multicolumn{2}{|l|}{Black hole mass}      \\ 
 $n_x \times n_z$            &	$100\times100$                                   & \multicolumn{2}{|l|}{Number of grid cells}    \\
 $R_{\rm max}$         &	$10^{20}~{\rm cm}$                               & \multicolumn{2}{|l|}{Grid domain size}    \\
  $R_{\rm mid}$         &	$R_{\rm max}$                               & \multicolumn{2}{|l|}{Extent of obscuring midplane}    \\
 $f_V$  	           &	$0.01$                                          & \multicolumn{2}{|l|}{Volume filling factor}          \\
 $\Omega / 4\pi$       &	$0.2$                                           & \multicolumn{2}{|l|}{Wind covering factor}    \\
  $\theta_{\rm max}$   &	$\arccos ( \cos \theta_{\rm min} - \Omega/4\pi)$ & \multicolumn{2}{|l|}{Wind outer opening angle}     \\
$r_{\mathrm{min}}$ 	   &    $300~r_g$                        & \multicolumn{2}{|l|}{Inner wind launch radius}      \\ 
$r_{\mathrm{max}}$ 	   &    $600~r_g$                        & \multicolumn{2}{|l|}{Outer wind launch radius}     \\ 
$v_0 (r_0)$            &	$c_s(r_0)$                                       & \multicolumn{2}{|l|}{Initial streamline velocity}    \\
$L_{\mathrm{bol}}$     &	$3\times10^{46}$~erg~s$^{-1}$                    & \multicolumn{2}{|l|}{Bolometric luminosity}    \\
$L_X$       &	$10^{45}$~erg~s$^{-1}$                                         & \multicolumn{2}{|l|}{X-ray luminosity (2-10 keV)}    \\
$\alpha_{\mathrm{X}}$  &	$-0.9$                                           & \multicolumn{2}{|l|}{X-ray spectral index}    \\
\hline 
\end{tabular}
\caption{{\sl Top:} The grid points used in the parameter search, resulting in 243 models in total. The values of the grid parameters for the two illustrative models discussed are also shown. 
{\sl Bottom:} Various fixed or derived parameters used in the simulations, with descriptions and further detail in the text.  
}
\label{grid_table}
\end{table}

\section{Results from a Simulation Grid}
\label{sec:results}

We now present results from the full simulation grid, comprising 243 photoionization and radiative transfer simulations with different wind model parameters. The simulations were run on the Cascake Lake nodes of the {\sl Cambridge Service for Data Driven Discovery (CSD3)} cluster, with an average run-time of $\approx 200$ core-hours per model. We vary the wind mass-loss rate, $\dot{M}_w$, the acceleration length $R_v$, the acceleration exponent, $\alpha$, the terminal velocity parameter, $f_{\infty}$ and the inner wind opening angle, $\theta_{\rm min}$. The outer wind opening angle, $\theta_{\rm max}$, is adjusted to maintain the desired covering factor $\Omega/4\pi$ (which is set to $0.2$ for the initial grid). We chose these specific parameters because they will modify the kinematics and/or line emissivity of the wind by altering the density and velocity along a streamline, with a resulting impact on emission line profiles. These choices are, however, somewhat arbitrary, which should be borne in mind when interpreting our results (see Sections \ref{sec:wind_params} and \ref{sec:limitations}). For each wind model, we compute synthetic spectra from $5^\circ-85^\circ$ at $5^\circ$ intervals over the wavelength range $500-7500$\,\AA. Convergence in each cell is evaluated by requiring that the electron and radiation temperatures are not changing significantly from cycle to cycle, and the heating and cooling are balanced, both within a relative tolerance of 5\,per cent. The convergence criteria are described in the code documentation and by \citet{long_modeling_2002} and \citetalias{matthews_stratified_2020}. The models generally converged extremely well: the median convergence was $97.6$\,per cent across the simulation grid. A handful of models did not converge well, particularly for $\alpha=1.5$, where the slow acceleration can lead to a dense and optically thick wind base. This structure makes it difficult to achieve good $r$-packet statistics in shielded regions, with a corresponding minimum convergence of $67.8$\,per cent. Nevertheless, the median convergence values for each value of $\alpha=0.5,1,1.5$ were, respectively, $99.2$, $97.2$ and $96.8$\,per cent. 

\begin{figure*}
    \centering
    \includegraphics[width=1.0\linewidth]{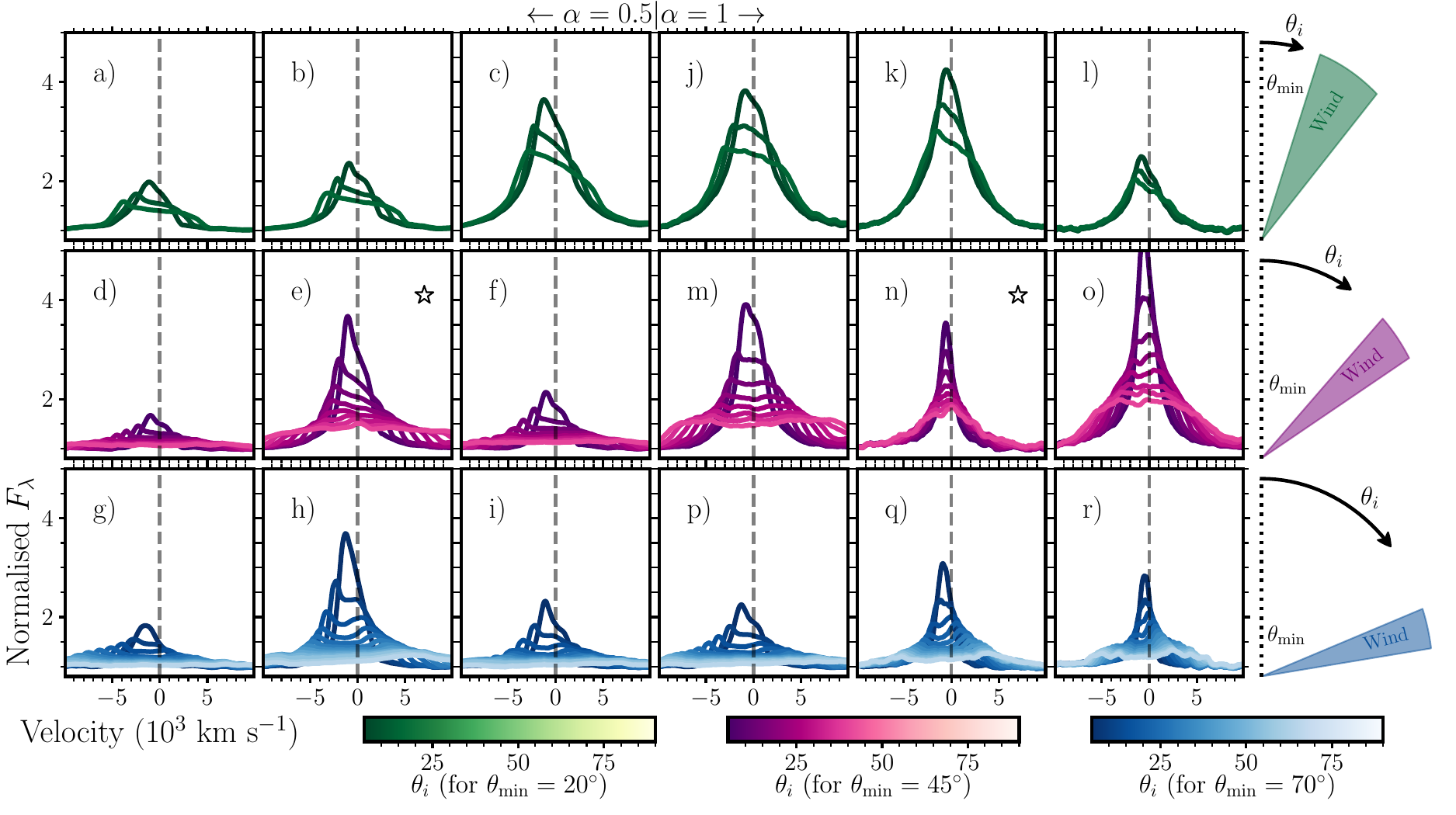}
    \caption{Continuum-normalised \civline\ line profiles as a function of velocity and at a range of inclinations, for 18/243 of the models in our initial simulation grid. The colour scheme is different for each wind inner opening angle $\theta_{\rm min}$ , with the polar, intermediate and equatorial winds shown using different colour schemes. The colour-map intensity denotes inclination, $\theta_i$, and the panels are labelled with a letter a) to r). The parameters for each model are given in Appendix~\ref{appendixa} with the corresponding labels, and the two fiducial models, e) and n), are marked with stars.}
    \label{fig:blueshift_grid}
\end{figure*}

\begin{figure*}
    \centering
    \includegraphics[width=0.49\linewidth]{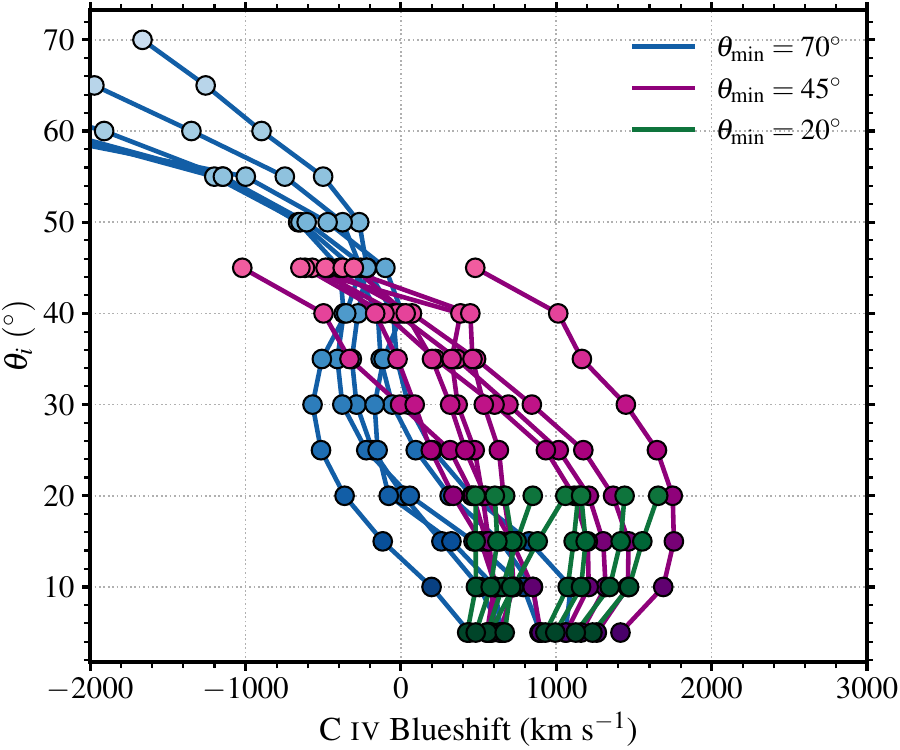}
    \includegraphics[width=0.49\linewidth]{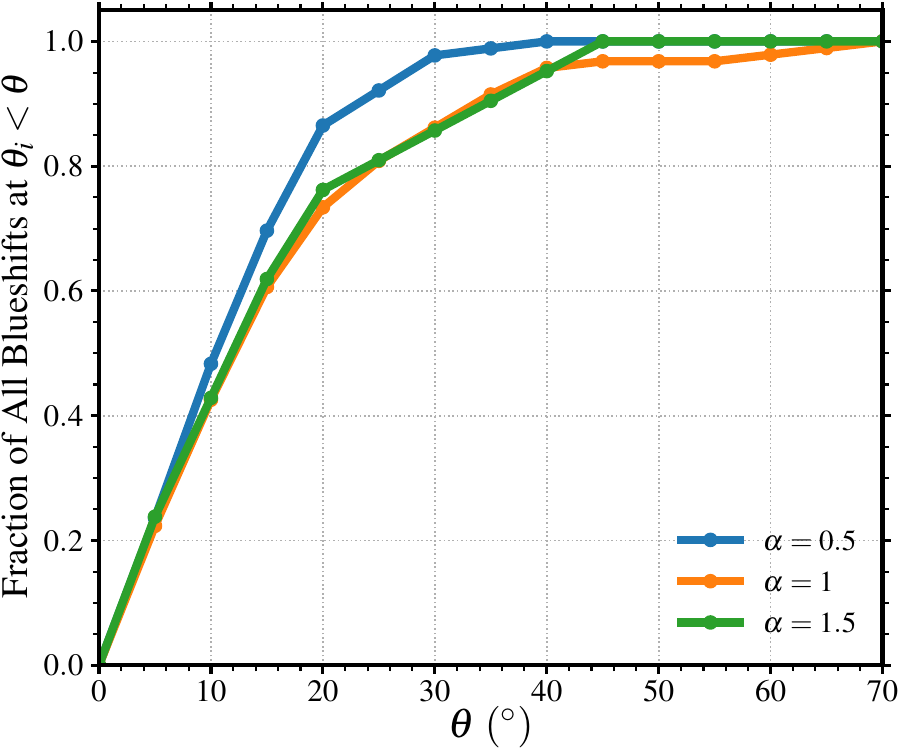}
    \caption{{\sl Left:} Blueshift as a function of inclination for the three considered wind geometries for $\alpha=0.5$. The colour-coding matches the colour-scheme in Fig.~\ref{fig:blueshift_grid}.  We only show inclinations from $5^\circ\leq\theta_i<\theta_{\rm min}$, and models are only plotted if they have \civline\ EWs exceeding $5$\AA\ over the range of $\theta_i$. {\sl Right:} The cumulative distribution function (CDF) showing the fraction of all the model spectra exhibiting blueshifts that are found at an inclination $\theta < \theta_i$. A CDF value of 1 means that no blueshifted spectra are found beyond that inclination. 
    }
    \label{fig:inclination}
\end{figure*}

\subsection{Line profiles and inclination trends} 
\label{sec:inclination}

In Fig.~\ref{fig:blueshift_grid}, we show normalised \civline\ line profiles for a subset (18 of 243) of the simulation grid, colour-coded by inclination. The parameters for each of these labelled models are given in Table~\ref{tab:params} in Appendix~\ref{appendixa}. We can see that the spectrum presented in the previous section is not an isolated case. Rather, blueshifted and asymmetric \civline\ line profiles are always observed for some viewing angles in these examples. In tandem with the heuristic demonstration above, this is the first main result of our work: biconical disc winds are successful in creating blueshifted emission lines.

We have focused on the $\alpha=0.5$ and $\alpha=1$ models in Fig.~\ref{fig:blueshift_grid}, partly for brevity and partly because the $\alpha=1.5$ models are slightly less effective at creating blueshifted asymmetries in the \civ\ line profile. The $\alpha=0.5$ and $\alpha=1$ models in Fig.~\ref{fig:blueshift_grid} match each other in the other parameters, as can be seen in Table~\ref{tab:params}. 
For completeness, the corresponding spectra for $\alpha=1.5$ are shown in Appendix~\ref{appendixa}, and the full set of \civ\ line profiles from all 243 models are given in the {\sl Supplementary Material}. Inspection of these spectra reveals that blueshifted asymmetries are ubiquitous in spectra that show prominent lines, with the basic conditions for their formation being that the wind has a sufficiently high \civline\ luminosity and that the midplane obscures the receding wind cone.

Some interesting trends with inclination can be seen in the line profiles, where we define inclination, $\theta_i$ with respect to the symmetry axis, such that a face on disc is seen at $\theta_i = 0$. The lines are narrower at low inclination and broaden towards high inclination. In some cases, at high inclination the blue wing of the line profile can be absorbed, leading to a {\em redshifted} asymmetry in the line profile -- prominent examples of this behaviour can be seen in panels q) and r). As also found by \cite{zamanov_kinematic_2002,gaskell_case_2016}, redshifted line profiles tend to occur at a critical inclination of $\theta_{\rm red} \approx 90^\circ - \theta_{\rm min}$, although this is a necessary but not sufficient criterion for forming redshifted asymmetries in our simulations. The reason these redshifted profiles form is actually quite subtle, since simply viewing a receding component to the wind (as occurs when $\theta_i \gtrsim \theta_{\rm red}$) is not enough to produce a redshifted asymmetry - in fact, without radiative transfer, one would expect a fairly symmetric profile in this case. The redshifted asymmetry is introduced due to the different path lengths (and therefore different optical depths / absorbing columns) traversed by photons from the blue and red wings of the line. In the receding part of the wind, photons escape to the observer along the normal to their streamline which is roughly the shortest path to the edge of the wind. The blueshifted line photons in the approaching part of the wind have a longer path  to travel to escape and so are more readily absorbed, leading to a redshifted asymmetry. Further discussion of the shape of the emission lines and a comparison to the observations can be found in Section~\ref{sec:shapes}. 

To examine the blueshift evolution with viewing angle more concretely, we show \civ\ blueshift (calculated from equation~\ref{eq:blueshift}) as a function of inclination in Fig.~\ref{fig:inclination}, for the $\alpha=0.5$ models (the same plots for the other values of $\alpha$ can be found in Appendix~\ref{appendixa}). For all models with reasonable strength emission lines, we see a clear evolution from blueshifted profiles at low inclinations and redshifted or symmetric profiles at higher inclinations. Generally speaking, blueshifts are only observed at low inclination, for all wind geometries. To quantify this further, we show the cumulative distribution function (CDF) showing, for each value of $\alpha$, the fraction of all the model spectra exhibiting blueshifts that are found at an inclination $\theta < \theta_i$. We define a `blueshifted' line as having a blueshift exceeding $500~{\rm km~s}^{-1}$ for this purpose. A CDF value of 1 means that no blueshifted spectra are found beyond that inclination. For all values of $\alpha$, at least $70$\,per cent of the spectra with \civ\ blueshifts in our model grid are seen at $\theta_i \lesssim 20^\circ$, and $\gtrsim95$\,per cent are viewed at $\theta_i \lesssim 40^\circ$. Only a handful of blueshifted line profiles are seen at $\theta_i \geq 50^\circ$ for any wind geometry or value of $\alpha$. 

These results demonstrate that blueshifts are preferentially seen at low inclinations, but there are a couple of caveats and complications. Firstly, in these plots we only consider $\theta_i \leq \theta_{\rm min}$, that is, we have not considered viewing angles looking into the wind cone -- since these sightlines often produce BALs (see section~\ref{sec:velocity_bals}) -- or looking underneath the wind. Along the latter class of sightlines, quite high EW emission lines can be observed as the continuum is often suppressed; we refer the reader to the discussion by \cite{matthews_testing_2016}. Our neglect of high inclination viewing angles means that the effect of blueshifts only being produced at low inclination could be artificial. However, this is not the case. Considering only the models with an equatorial disc wind geometry (and so also with a large range of angles with $\theta_i \leq \theta_{\rm min}$), we still see a clear trend from high \civ\ blueshift at low inclination to low or negative blueshift at high inclination in the left-hand panel of Fig.~\ref{fig:inclination}. In addition, in our adopted geometry it makes intuitive sense that blueshifts would be produce at low inclinations because the midplane acts as the obscurer of the red wing.  
The second complication here is that blueshift can in some cases initially increase with inclination, as occurs in, for example, a number of the $\theta_{\rm min} = 45^\circ$ curves in the left-hand panel of Fig.~\ref{fig:inclination}. Again, this is caused by projection effects. As $\theta_i$ increases, the outflow velocity component projected into the line of sight can also increase, leading to a slightly stronger blueshift. Neither of these two caveats invalidate our main conclusion here: emission line blueshifts are preferentially produced at low inclination, and this behaviour is an inevitable consequence of the projection effects within, and symmetry axes of, the expected quasar disc wind geometry. 

\subsubsection{Impact of rotation and line transfer}
\label{sec:impacts}
To assess the impact of rotation and line transfer, we take the two fiducial models, e) and n), and, using the same ionization and temperature structure, re-simulate their spectra without rotation or line transfer. Specifically, for the rotation test, we set all rotational velocity components to zero, and for the line transfer test we set the Sobolev optical depth to zero during the actual radiative transfer process (it is still included as a local effect via the Sobolev escape probabilities). The goal of this exercise is not to mimic any physically realistic situation. Instead, the aim is to assess the role rotation and line transfer play in determining the shape of \civ\ emission lines in our synthetic spectra. 

\begin{figure}
    \centering
    \includegraphics[width=\linewidth]{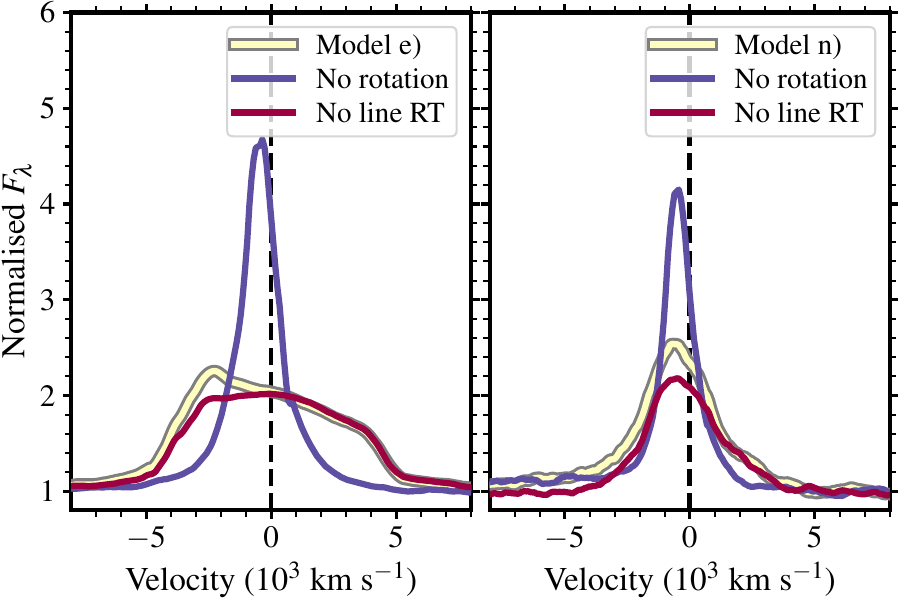}
    \caption{The impact of line transfer and rotation on the \civ\ line profile at $\theta_i = \view^\circ$, for our two fiducial models e) and n). Rotation acts to broaden the line profile, particularly in model e). At this inclination, line opacity within the wind can act to increase the line flux by introducing a resonantly scattered component to the emission line, particularly in the blue wing. 
    }
    \label{fig:impacts}
\end{figure}

The results from these two tests are shown in Fig.~\ref{fig:impacts}. We find that neglecting rotation leads to a single-peaked blueshifted profile which is also narrower. The impact is more pronounced for Model e) because the line is formed closer to the wind base where the kinematics are more dominated by rotation. In contrast, rotational velocity has a fairly modest effect on the model n) line profile, because outflow is more important in the line formation region. The overall line profile can be thought of as a convolution of a rotational component with an outflowing one, with the relative importance of each determined by the kinematics at the point where \civline\ is emitting efficiently -- although, in reality, the line formation region is stratified in height and thus velocity, as discussed in Section~\ref{sec:velocity_bals}. 

The neglect of line opacity within the flow leads to a modest, velocity-dependent suppression of the overall line flux. This suggests there is a resonant scattering component to the line profile, caused by scattering of \civ\ line photons into the line of sight. In model e), this occurs preferentially at blue-shifted velocities and is responsible for a horn-like feature, reminiscent of one side of a double-peaked disk emission line profile. In model n), the scattering contribution is less confined to one part of the line profile although it does occur exclusively at velocities $\lesssim 1000~{\rm km~s}^{-1}$. These results suggest that radiative transfer effects within the wind are important for determining the overall line profile, and so only knowing the emissivity profile, say, of emission lines within the wind is not sufficient for an accurate characterisation of their shapes.

\subsection{Comparison to observations} 
\label{sec:shapes}

\subsubsection{Skew and line profile shape}
\label{sec:skew}
\begin{figure}
    \centering
    \includegraphics[width=\linewidth]{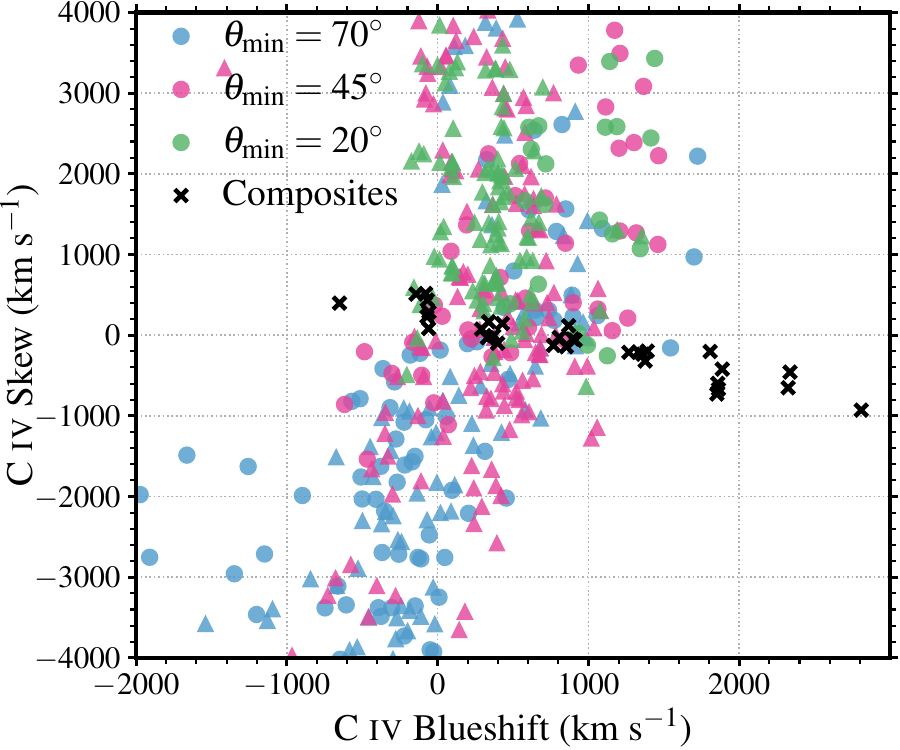}
    \caption{
    Skew as a function of blueshift for the model grid (circles and triangles) and composite spectra (black crosses). The circles and triangles denote the $\alpha=0.5$ and $\alpha=1$ results, respectively. We find that most of the model spectra with positive blueshifts also have positive skews (that is, their line peaks are blueward of $\lambda_{\rm half}$). In contrast, almost all of the composite spectra have negative skews (their line peaks are redward of $\lambda_{\rm half}$). 
    }
    \label{fig:skew}
\end{figure}

One characteristic feature of the line profiles presented in Fig.~\ref{fig:blueshift_grid} is that the peak in the line profile tends to be skewed towards the blue, such that the velocity of the peak is more highly blueshifted than the `blueshift' calculated from equation~\ref{eq:blueshift}. Using {\sl skew}, as defined by equation~\ref{eq:skew}, we can quantify this effect for both the observations and models, the results of which are presented in Fig.~\ref{fig:skew} for the $\alpha=0.5$ and $\alpha=1$ models. In the composite spectra from \citetalias{rankine_bal_2020}, we find a roughly linear relationship between skew and blueshift, with more blueshifted lines having more {\em negative skews}. In other words, and as can be seen from Fig.~\ref{fig:composite}, weak line quasars with strong blueshifts tend to have strong asymmetries and a more extended blue wing, with their line peaks lying redward of the wavelength that bisects the line flux. This is generally the opposite to what we see in the models, where high-blueshift spectra tend to have positive skews in Fig.~\ref{fig:skew}. Thus, while we are successful in producing blueshifts of a couple of thousand ${\rm km~s}^{-1}$ in our models, the line profiles are normally skewed in the wrong direction compared to the observations. Furthermore, in the models, there is a clear weak positive correlation between blueshift and skew, in the opposite sense to the anti-correlation in the composite spectra. This result might hint at problems with the wind geometry or kinematics in our model compared to real outflows in quasars. However, it is also important to stress once again that Fig.~\ref{fig:skew} is in some sense an `apples versus oranges' comparison, since we have simulated a quasar with fixed accretion rate and mass, and we are comparing to composite spectra built from a population of quasars spanning a range of masses and Eddington fractions. 

There are a few other effects in Fig.~\ref{fig:skew} that are worth discussing. The first can already be inferred from Fig.~\ref{fig:inclination}, that the most extreme blueshifts in the model do not exceed $2000~{\rm km~s}^{-1}$, whereas the composite spectra (and the overall \civ\ emission space) extends to $\sim 3000~{\rm km~s}^{-1}$. Our wind model cannot reproduce these extreme blueshifts, suggesting that these outflows are particularly dense or powerful such that \civ\ line formation can take place efficiently at high velocity (see discussion in subsequent subsection). This apparent discrepancy is perhaps expected. The $\sim 3000~{\rm km~s}^{-1}$ \civ\ blueshifts are observed only in a small population of quasars with Eddington fractions of $\gtrsim 0.5$ and slightly higher masses than we have considered here \citep{temple_testing_2023}. These quasars would be expected to have softer SEDs (or higher UV to X-ray flux ratios) more conducive to line-driving, which, combined with the high Eddington ratios, might be expected to lead to more powerful and/or faster outflows \citep[e.g.][]{giustini_global_2019}. The second notable point is that there are points more or less overlying some of the observational data. However, inspection of these models reveal that none of them produce line profiles with similar shapes to the observations, as can be seen from the full set of line profiles in the {\sl Supplementary Material}. Finally, we note that different wind geometries -- denoted by different colours in Fig.~\ref{fig:skew} -- produce different patterns in this parameter space, with more polar wind geometries tending to have more positive skews. The fact that there is a relationship between the line profile shape and wind geometry merits further investigation, implying that different or more complex wind geometries to those considered here could lead to a better agreement with observed spectra (see section~\ref{sec:discuss_physics}).

\subsubsection{Equivalent width and line luminosity}
\label{sec:ew}
When inspecting all 243 of the \civline\ line profiles from our simulation grid (see {\sl Supplementary Material}), we find that blueshifts are always produced at low inclination as long as the emission line is obviously present. In a number of cases, the wind is over- or under-ionized, and the continuum is more or less featureless in the UV. Thus, while Fig.~\ref{fig:blueshift_grid} shows a favourable selection of models that all produce strong lines, the blueshift formation is an inevitable consequence of obscuring the receding part of the wind -- as long as the lines are strong enough and the obscuring midplane is present, a blueshifted asymmetry is formed. However, ensuring enough of the wind lies in the right ionization state to form emission lines of sufficiently high EWs (that is, comparable to those in observed spectra) can be a problem, as we have discussed in previous work (\citealt{matthews_testing_2016}; \citetalias{matthews_stratified_2020}). 

As we have described and illustrated in Section~\ref{sec:observed}, there is a seemingly profound relationship between EW and blueshift in quasar spectra. The true \civ\ emission space is built up from quasars with a range of masses and bolometric luminosities, whereas our models are run for one specific value of each. As a result, it would be a little misleading to compare directly the range of EWs and blueshifts in our models to those from a population of quasars, since our intention is not to simulate such a population. Instead, we can compare the overall luminosities and EWs of \civline\ computed from our synthetic spectra to observations. This exercise has already been carried out with similar models in our previous work (\citetalias{matthews_stratified_2020}, see their fig.~6) where we found that models that avoided over-ionization produce line luminosities and EWs similar to those in observed spectra of quasars of comparable luminosity. The same is true with our simulation grid here: there are still a substantial number of models that do not produce noticeable emission lines, because they are either under- or over-ionized. Considering all models, the median \civ\ line luminosity at viewing angles $\leq \theta_{\rm min}$ is $L_{1550} \approx 1.4\times 10^{44}{\rm erg~s}^{-1}$, spanning a wide range of luminosities ($\approx 10^{42}-10^{45.5}{\rm erg~s}^{-1}$). For the subset of models shown in Fig.~\ref{fig:blueshift_grid}, the line luminosities at the same viewing angles are a little higher, since these models have been selected to have obvious \civ\ line profiles, with a median of $L_{1550} \approx 5.4\times 10^{44}{\rm erg~s}^{-1}$. For reference, SDSS quasars with $\lambda L_{1350} \approx 10^{46}~{\rm erg~s}^{-1}$ typically have \civ\ line luminosities $L_{1550} \approx 3\times 10^{44}~{\rm erg~s}^{-1}$, similar to the median line luminosity in our models.

\begin{figure*}
    \centering
    \includegraphics[width=1.0\linewidth]{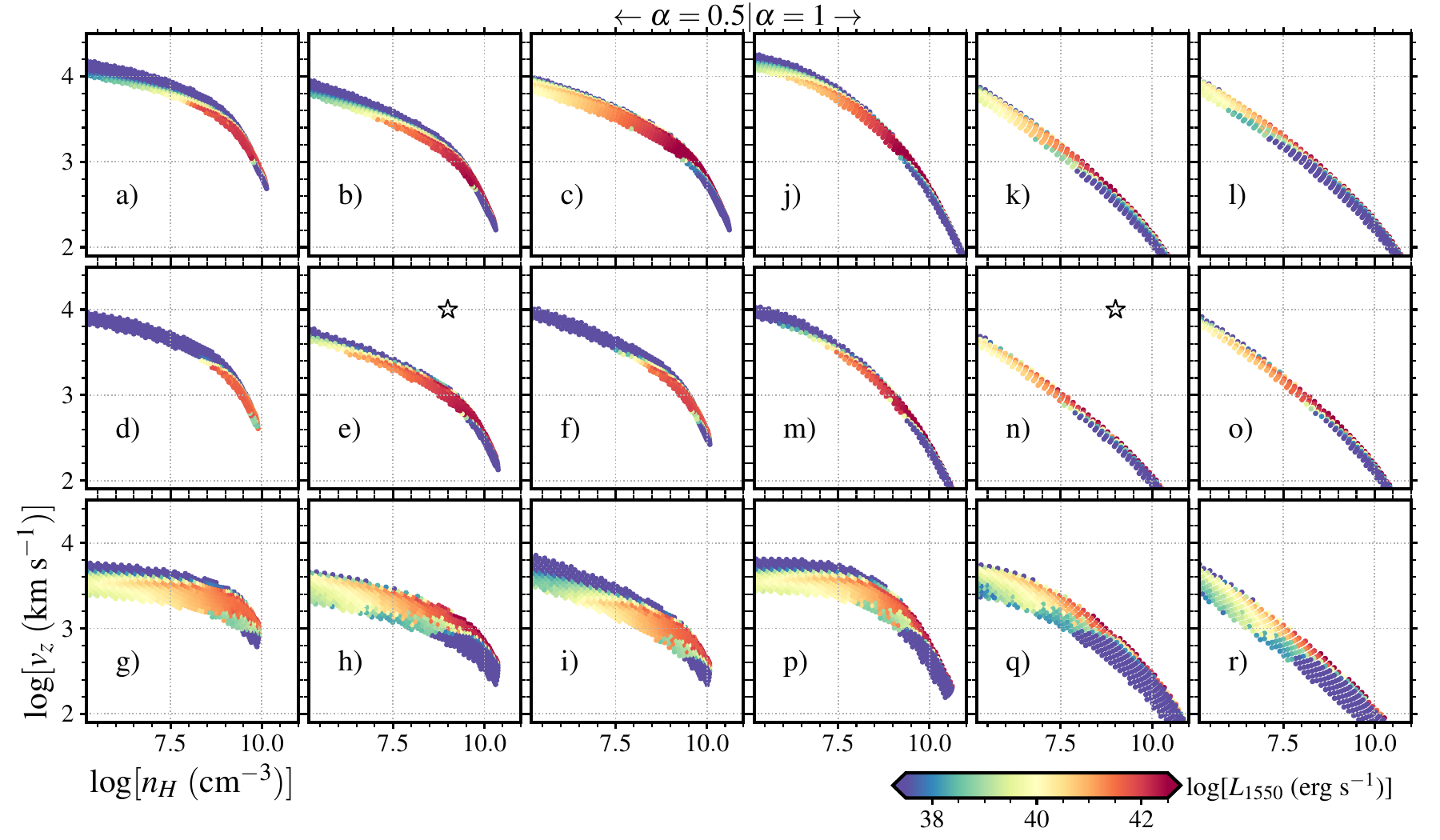}
    \caption{The phase space of the line formation region. The logarithmic \civline\ local line luminosity in each cell is shown in colour as a function of electron density, $n_e$, and vertical velocity, $v_z$. The letters labelling models in each panel matches Fig.~\ref{fig:phase}, and the two fiducial models, e) and n), are marked with stars. 
    The line luminosity is the energy per unit time escaping the local Sobolev region. In all cases, the vast majority of line emission is produced in regions with $v_z \lesssim 3000~{\rm km~s}^{-1}$.
    }
    \label{fig:phase}
\end{figure*}

\begin{figure*}
    \centering
    \includegraphics[width=\linewidth]{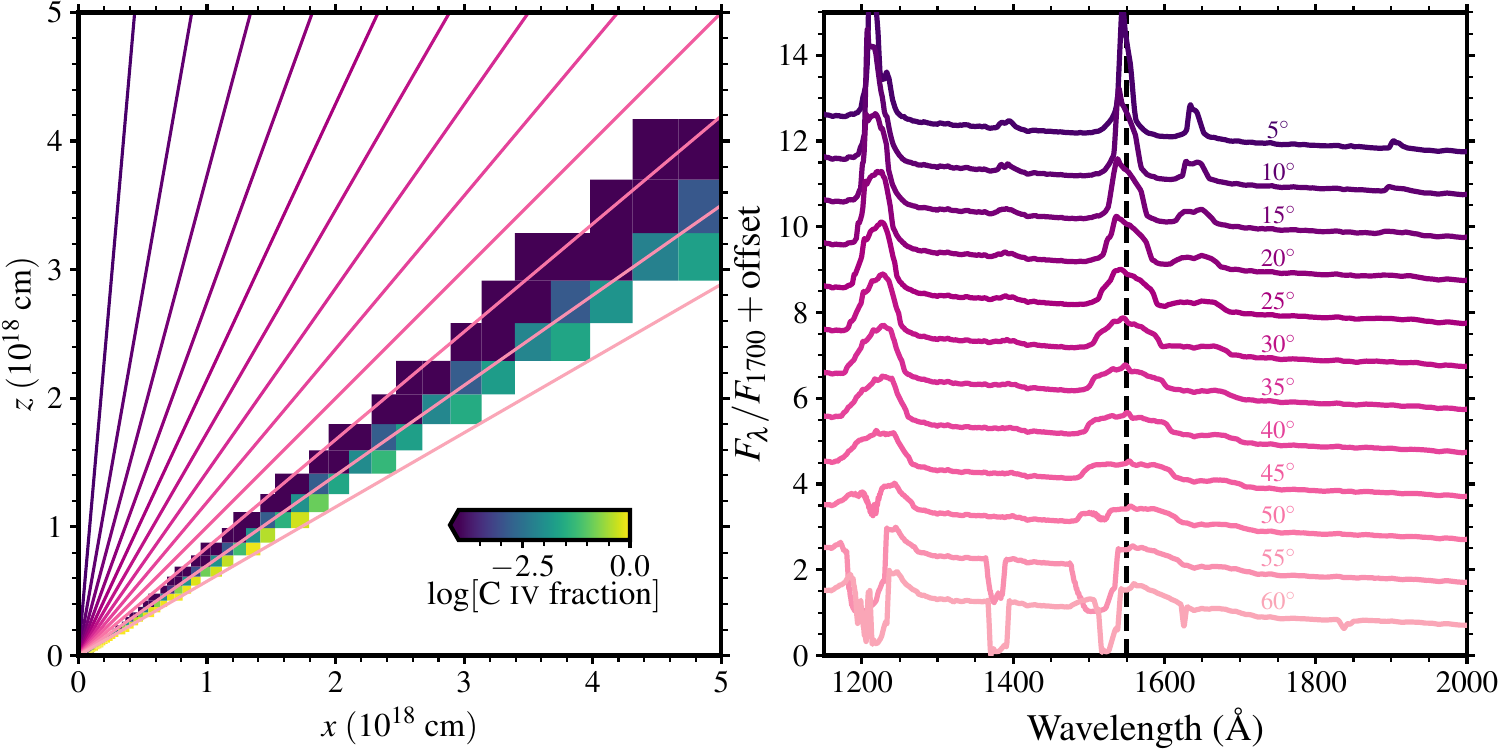}
    \caption{An example of the formation of \civ\ blueshifts and BALs in the same wind structure, with results from Model e). {\sl Left:} The ionization parameter in the wind, together with sightlines through the wind structure, colour-coded by inclination with colours matching the right-hand panel. {\sl Right:} Synthetic spectra at viewing angles from $5^\circ$ to $60^\circ$ at $5^\circ$ intervals. The spectra are normalised to the $1700$\AA\ flux with an offset of $1$ applied for clarity. The \civline\ wavelength is marked with a dashed line. BALs are formed at viewing angles that intersect moderately ionized material within the wind cone, and \civ\ emission line blueshifts are formed at low inclinations.  
    }
    \label{fig:bal}
\end{figure*}

\begin{figure*}
    \centering
    \includegraphics[width=\linewidth]{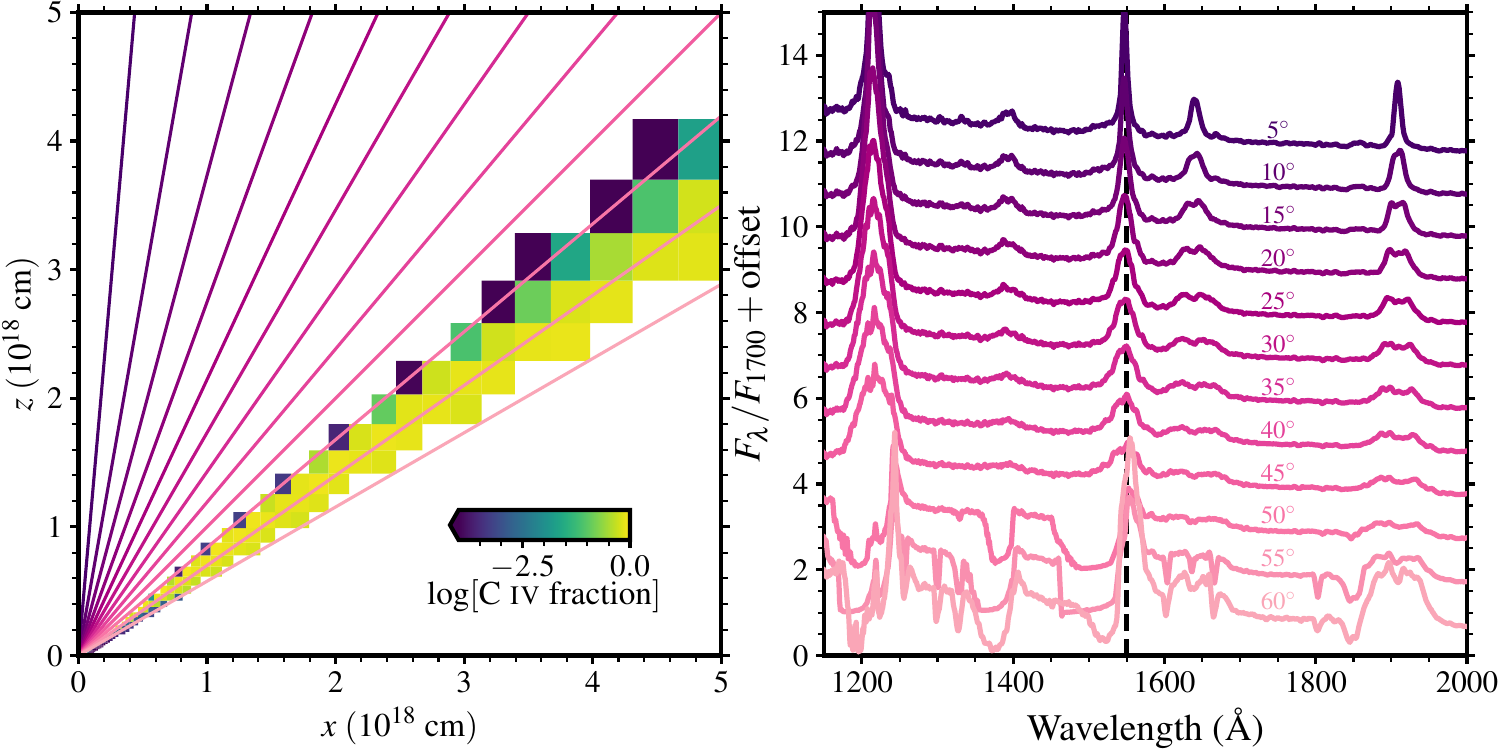}
    \caption{As Fig.~\ref{fig:bal}, but for $\alpha=1$ (Model n). 
    }
    \label{fig:bal2}
\end{figure*}

\subsection{Velocity structure and BAL formation}
\label{sec:velocity_bals}

 To allow an inspection of the kinematics of the line formation region, we show a `phase space' plot of vertical velocity, $v_z$, and electron density, $n_e$, in Fig.~\ref{fig:phase}, with the panels matching those in Fig.~\ref{fig:blueshift_grid}. Each point shows the location in $n_e,v_z$ parameter space for each cell in the wind model, colour-coded by the local \civline\ line luminosity in each cell. This quantity includes the local Sobolev escape probability but the line photons can be absorbed further out in the wind, so do not necessarily escape to infinity. The bright regions in the plot show the typical velocities and densities where the \civline\ line is forming. There is some variation in the velocity and density structure across the grid (due to changes in the grid parameters $\theta_{\rm min}$, $\dot{M}_w$, $R_v$ and $f_\infty$), but typically the main line formation region spans the density range $8.5 \lesssim \log(n_e) \lesssim 10.5$ and the velocity range $300-3000~{\rm km~s}^{-1}$. To make things more quantitative, we can define $v_{90}$ as the $v_z$ such that $90$\,per cent of the \civ\ line emission comes from regions with $v_z < v_{90}$. In all models shown in  Fig.~\ref{fig:phase}, $v_{90} < 3050~{\rm km~s}^{-1}$, and the mean and median $v_{90}$ values are $1311~{\rm km~s}^{-1}$ and $1335~{\rm km~s}^{-1}$, respectively.
 
The location of the line formation region in the $n_e,v_z$ plane makes clear why characteristic blueshift velocities of a few $1000~{\rm km~s}^{-1}$ are observed in the spectra computed from these wind models; at higher velocities the density is lower and the emission line emissivity is much lower. The line emissivity is driven by two factors. The first is ionization-state -- at low densities and high velocities the wind is over-ionized (mostly in \ion{C}{v} and upwards) and \civ\ line formation is ineffective. The second is density or emission measure. Since the line is collisionally excited, its emissivity is proportional to $n_e^2$, so even for fixed ionization state a drop in density leads to a corresponding drop in emissivity. Both these factors act to confine the \civ\ line formation region to near the base of the outflow where velocities are relatively modest and densities are relatively high. The line formation region can be moved around in the flow by, for example, modifying the velocity law (see Section~\ref{sec:wind_params}).

The wind does eventually accelerate towards its terminal velocity, becoming a more tenuous low density plasma with velocities exceeding $10,000~{\rm km~s}^{-1}$. We find, for observers looking directly into the wind cone, that \civ\ BALs are often formed in the extended high-velocity flow. In Figs.~\ref{fig:bal} and \ref{fig:bal2}, we show the simultaneous formation of 
\civ\ BALs and blueshifts in the same outflow for Models e) and n). In both cases, spectra in the rest-frame UV from low to high inclination are shown in the right-hand panel with offsets applied for clarity, and the viewing angles through, and ionization structure of, the wind are shown in the left-hand panel. At low inclination, blueshifted emission lines are observed, and at angles $\theta_{\rm min} \lesssim \theta_i \lesssim \theta_{\rm max}$ BALs are observed in \civline\ and \nvline, and often also \sivline. As we found previously \citep{matthews_testing_2016}, LoBAL features in, e.g. \ion{Al}{iii} $\lambda$$\lambda$1855,1863 and even \ion{Fe}{ii} and \ion{Fe}{iii} absorption lines can be observed at high inclination angles that look through the low-ionization, shielded part of the wind.  

Consistent with our earlier work \citep{higginbottom_simple_2013,matthews_testing_2016}, BALs are only formed when the ionization state is sufficiently moderate and a significant line-opacity is present in the flow. As such, BALs typically form in the same models that are successful in producing strong emission lines, although the BAL formation region is significantly more extended than the emission line region. Figs.~\ref{fig:bal} and \ref{fig:bal2} represent another key result of our work: BALs and \civ-emission blueshifts can form in the same wind structure within a geometric unification scheme. This result is discussed critically in the context of observational results in Section~\ref{sec:discuss_bals}.

\begin{figure*}
    \centering
    \includegraphics[width=0.9\linewidth]{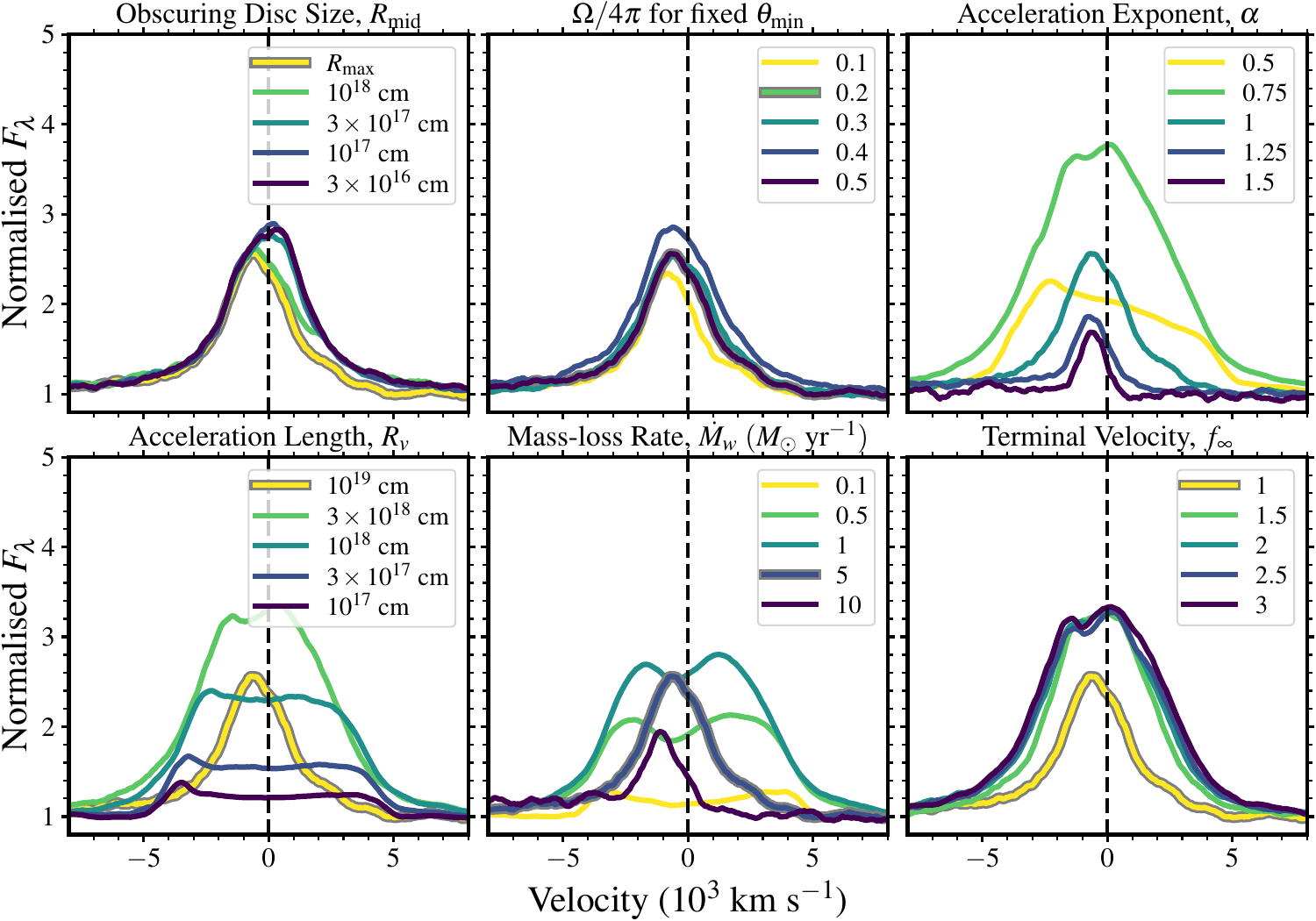}
    \caption{The dependence of the \civ\ line profile on model parameters at $\theta_i = \view^\circ$. In each panel, we show how the continuum normalised \civline\ profile varies in shape as a function of the labelled parameter, taking steps in parameter space away from the default model e). In each case, the colour denotes different values of the parameter and the default model line profile is highlighted in grey. Decreasing the obscuring midplane size, $R_{\rm mid}$ starts to reveal more and more of the redshifted part of the profile as the receding wind becomes visible. Covering factor makes very little difference for fixed $\theta_{\rm min}$. The acceleration exponent $\alpha$ has a significant impact on both the line EW and shape (see discussion in Section~\ref{sec:wind_params}), with the line first increasing then decreasing in intensity. The blueshifted peak also moves to lower velocity and the profile overall more closely resembles a single peaked quasar line. Decreasing the acceleration length basically acts to suppress the line EW, but does very little to the overall shape and line peak velocity or skew, and similar effects are achieved from increasing $f_\infty$.}
    \label{fig:step_par}
\end{figure*}

\subsection{Dependence on model parameters: the velocity law is key} 
\label{sec:wind_params}

Our simulation grid is built from five main model parameters, and the synthetic spectrum is determined by the combination of these plus observer inclination, $\theta_i$. We have already discussed the impact of $\theta_{\rm min}$ and $\theta_i$. To investigate the sensitivity of the \civ\ line profile shape to the model parameters, we take Model e) as a starting point and then take steps away from this model in the four remaining parameters ($\alpha$, $R_v$, $f_\infty$, $\dot{M}_w$). In addition, we also varied the obscuring disc size, $R_{\rm mid}$ and the covering factor, $\Omega/4\pi$. To provide slightly finer resolution, we supplement our main simulation grid with a few additional model runs so as to have spectra at five values of each parameter. 

The results of this `step' parameter search are shown in Fig.~\ref{fig:step_par}, designed to give the reader a feel for how the \civ\ emission line varies with each of these quantities. From this figure, we can see varying degrees of impact on the line profile shape and line EW. For example, decreasing the acceleration length (or increasing the terminal velocity through the parameter $f_\infty$) acts to weaken the emission line but has little impact on the line profile shape. By contrast, the acceleration exponent has a significant effect, changing the peak velocity of the line as well as its skew and EW. This dependence of the line profile shape on the acceleration exponent, and therefore the shape of the velocity law, is particularly interesting, so we discuss it further in Section~\ref{sec:discuss_physics}. Additionally, we can see that the extent of the opaque midplane, $R_{\rm mid}$, is important in determining whether blueshifts form, as expected from the discussion in section~\ref{sec:heuristic}. In particular, blueshifts start to emerge once $R_{\rm mid} \gtrsim 10^{18}~{\rm cm}$, a threshold value matching the characteristic horizontal distance of \civline\ line formation in this model; we discuss the opaque midplane and this physical scale further in section~\ref{sec:discuss_midplane}. 

\begin{figure*}
    \centering
    \includegraphics[width=1.0\linewidth]{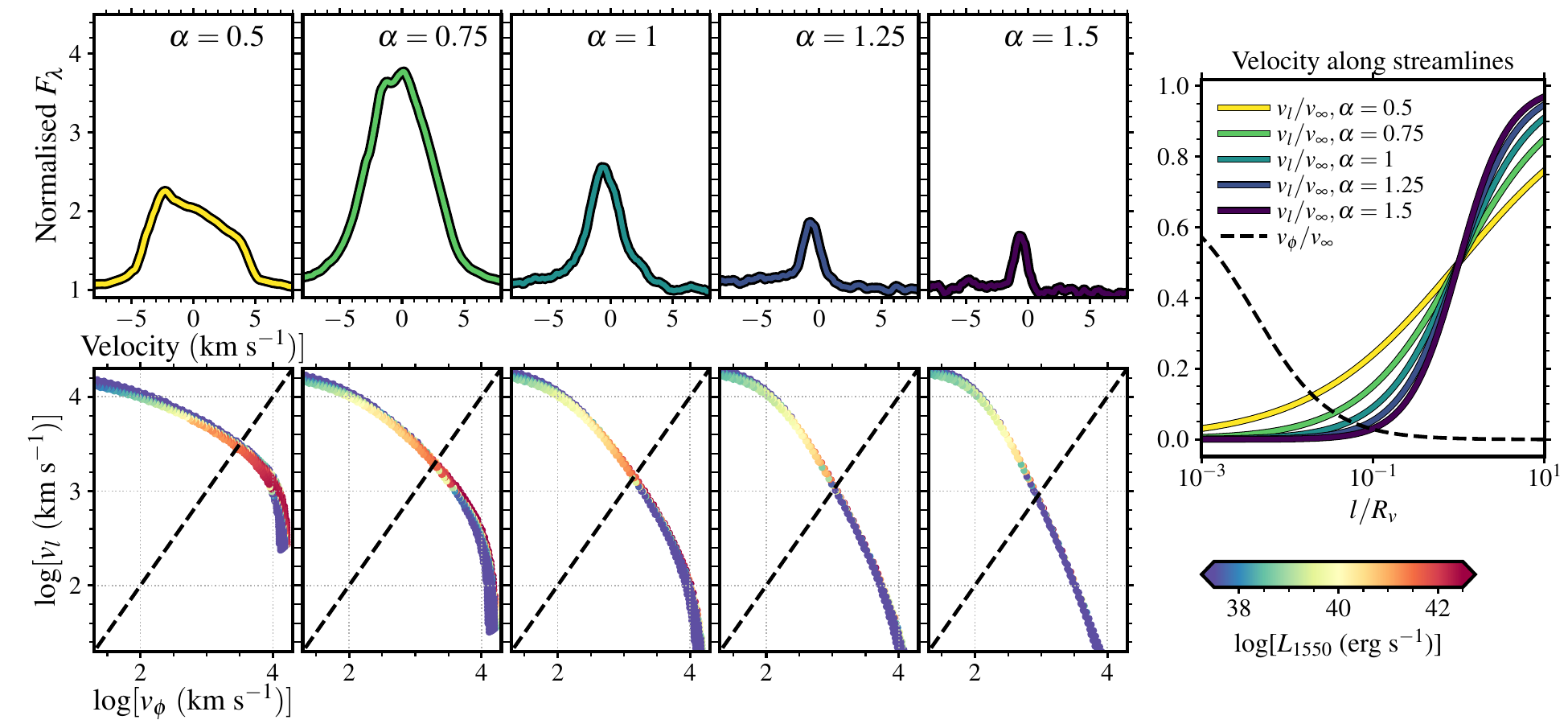}
    \caption{The change in line profile shape with velocity law can be explained by the location of the line formation region in velocity space. The top panel shows the \civ\ line profiles for each of the five values of $\alpha$ used in the step parameter search from Fig~\ref{fig:step_par}. The corresponding bottom panel shows the location of the wind cells in $v_l,v_\phi$ parameter space, colour-coded by line luminosity as in Fig.~\ref{fig:phase}, with a dashed line marking when $v_l=v_\phi$. The poloidal and rotational velocities along a streamline are shown in the rightmost panel, with colour-coding matching the line profiles.}
    \label{fig:lineform}
\end{figure*}

The reason for the change in line profile shape with the acceleration exponent, $\alpha$, can be understood more clearly by examining how changing $\alpha$ affects the location of the line formation region in velocity space. In Fig.~\ref{fig:lineform}, we show the \civ\ line profiles for each of the five values of $\alpha$ (top panels), together with the location of the wind cells in $v_l,v_\phi$ parameter space, colour-coded by line luminosity as in Fig.~\ref{fig:phase} (bottom panels). Additionally, we show the poloidal velocity along a streamline from equation~\ref{eq:velocity} for each value of $\alpha$ in the right hand panel, as well as the rotational velocity along the streamline as calculated from conservation of angular momentum. As $\alpha$ is increased, the wind accelerates more slowly and the line formation region moves to higher ratios of $v_l/v_\phi$, producing emission line profiles that are narrower, single-peaked, and more dominated by outflow. By contrast, for $\alpha=0.5$ the line formation is mostly confined to regions close to the disc where $v_\phi$ is high.   

Further intuition can be obtained here by considering a simple toy model where the \civ\ line forms at a fixed ionization parameter $\xi$. Assuming that changing the velocity law doesn't alter the line formation radius all that much, and considering fixed ionizing luminosity and wind mass loss rate, then the ionization parameter is set by the density. The density, in turn, is largely determined by the poloidal velocity of the wind \citep{shlosman_winds_1993}. In this case, one might expect that the line formation region occurs around a given fixed value of $v_l$. The shape of the line profile, and the relative dominance of rotation and outflow components, is then set by $v_l/v_\phi$. Increasing $\alpha$ means that the wind accelerates more slowly, but the variation of $v_\phi$ along a streamline is fixed (see right-hand panel of Fig.~\ref{fig:lineform}). Thus, increasing $\alpha$ causes the `critical' value of $v_l$ to be reached at lower values of $v_l/v_\phi$, and thus the line profile becomes more single-peaked and less influenced by rotational effects. 

\begin{figure*}
    \centering
    \includegraphics[width=1.0\linewidth]{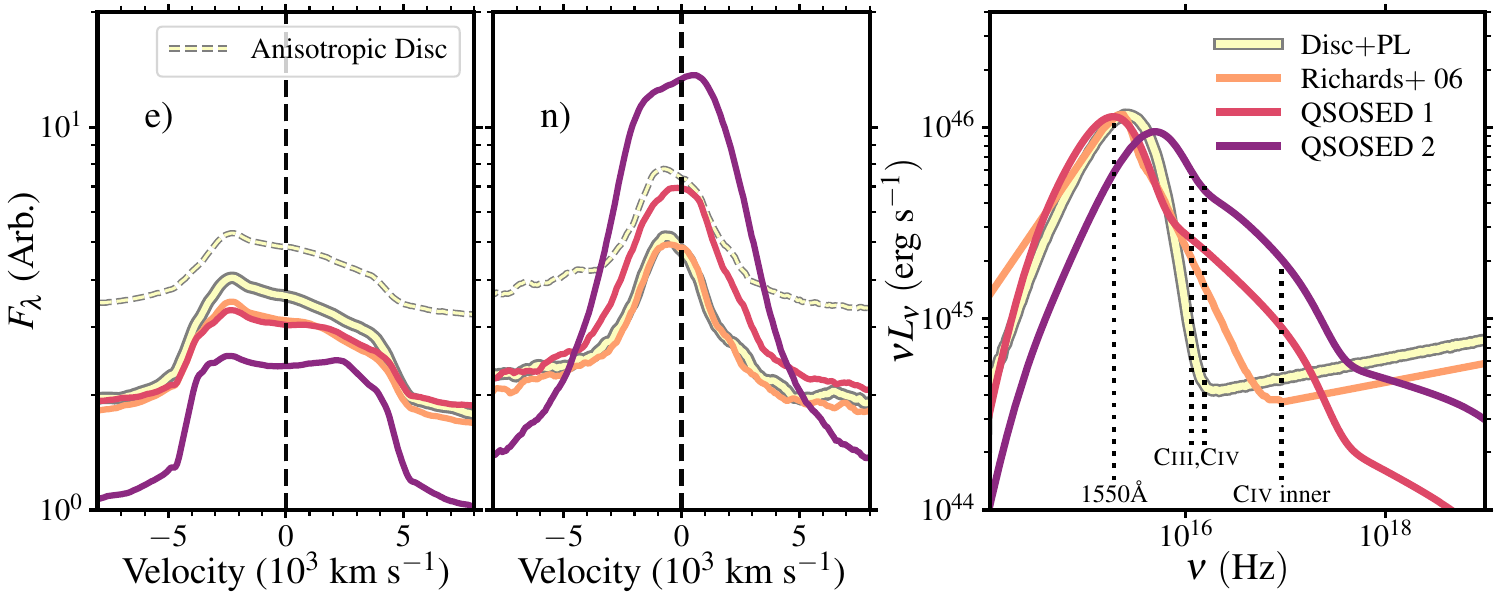}
    \caption{The dependence of the \civ\ line profile on the ionizing SED. In the two left-hand panels, the line profiles at $\theta_i = \view^\circ$ are shown for models e) and n) for different SED shapes, including one with an anisotropic accretion disc component. The colour-coding matches the right-hand panel, in which the four SEDs are shown. The key frequencies corresponding to the line wavelength and the C\textsc{iii}$\to$C\textsc{iv}, C\textsc{iv}$\to$C\textsc{v} and C\textsc{iv} inner shell ionization edges are marked with vertical dotted lines.}
    \label{fig:sed}
\end{figure*}

\subsection{Impact of SED shape and anisotropy} 
\label{sec:sed_impact}

For our initial simulation grid and fiducial model, we used the same isotropic `disc+power-law' SED as \citetalias{matthews_stratified_2020}. The anisotropy and shape of the illuminating SED is likely to play an important role in determining the observed spectrum: the flux anisotropy changes wind dynamics \citep{dyda_effects_2018,williamson_3d_2019} and emission line EWs \citep{risaliti_o_2011,matthews_quasar_2017}, while the frequency distribution modifies both the force multiplier \citep{dannen_photoionization_2019} and the observed emission line ratios \citep[e.g.][]{leighly_hubble_2004,richards_unification_2011}. 

In this work, we cannot account for dynamic effects directly, but we can investigate the sensitivity of our results to SED changes for a fixed wind structure. To do this, we take models e) and n), and fully re-simulate the ionization structure and resulting \civline\ emission-line profile for four additional SED choices: an identical SED shape with an anisotropic accretion disc \citep[following][]{matthews_testing_2016}, an isotropic SED based on the \cite{richards_sloan_2006} template, and two SEDs computed with the \textsc{qsosed} model of \cite{kubota_physical_2018}. The \textsc{qsosed} models were taken from the \cite{temple_testing_2023} simulation grid and are described by mass, Eddington-normalised accretion rate ($\dot{m}$) and dimensionless spin ($a_*$). We pick two models: \textsc{qsosed}1, with a BH mass of $10^{9.5}~{\rm M_\odot}$, $\dot{m}=0.468$ and $a_*=0$, and \textsc{qsosed}2, with a BH mass of $10^{9}~{\rm M_\odot}$, $\dot{m}=0.15$ and $a_*=0.998$. These two model spectra are quite different in terms of the hardness or colour of the SED and produce very different fluxes. We renormalised all of our model SEDs to have identical bolometric luminosities to that of our original model. As a result, the SEDs used are not necessarily physically consistent or realistic, but rather are designed to illustrate the variation of the line profiles for a wide range of spectral shapes and ionizing fluxes.  

The results from our SED investigation are shown in Fig.~\ref{fig:sed}. In the right-hand panel we show the four SEDs, while in the left-hand panels we show how the line profiles change in each case, for both models e) and n). Most of the SED changes have a relatively modest impact on the \civ\ line profile, with the results from the \cite{richards_sloan_2006} and \textsc{qsosed}1 SEDs showing rather similar profile shapes and line EWs to the fiducial model case. The main effect of the anisotropic disc is to increase the continuum flux (due to foreshortening and limb darkening), which acts to decrease the line EW slightly. Easily the most dramatic change is observed when using the rather extreme \textsc{qsosed}2 SED, which produces a significantly stronger line in the model n) case. This SED has a significantly stronger EUV component as well as a much weaker continuum in the near UV. The key frequencies corresponding to the line wavelength and the C\textsc{iii}$\to$C\textsc{iv} and C\textsc{iv}$\to$C\textsc{v} ionization edges, as well as the C\textsc{iv} inner shell edge, are marked in the right-hand panel of Fig.~\ref{fig:sed}. The weaker continuum underlying the line means the EW is significantly higher, but there is also an effect on the line luminosity, which increases. In tandem, the shape of the line profile has changed and now has a slight redshifted asymmetry due to absorption of the blue wing by the wind. However, these more dramatic changes are only seen for this drastically different SED shape, which has much higher fluxes around the key ionization edges (see right-hand panel of Fig.~\ref{fig:sed}). This SED is unlikely to be particularly realistic for a typical quasar,  

Overall, we can draw two main conclusions from our SED investigation. First, that significant changes in SED shape, as might be seen across a population of quasars, can clearly alter the \civ\ blueshift and EW; this result is interesting given the apparent dependence of on \civ\ emission space location on SED shape \citep{richards_unification_2011,rankine_bal_2020,temple_testing_2023}. Second, that our overall results are not particularly sensitive to our adopted SED, since three reasonable choices with the same bolometric luminosity but different soft X-ray excesses and ionizing luminosities all produce blueshifted \civ\ emission lines with similar shapes and EWs. 


\section{Discussion}
\label{sec:discuss}

\subsection{Line profiles as  a probe of disc wind physics}
\label{sec:discuss_physics}

We found in Section~\ref{sec:wind_params} that the velocity law -- in particular the value of $\alpha$ -- can have a dramatic impact on the observed shape of the \civline\ line profile, and we discussed how the line formation region moves around in velocity space with these changes in $\alpha$. In a general sense, this finding suggests that emission line profile shapes could be used as a probe of the velocity law of the wind.  Although the detailed wind velocity law from different wind driving mechanisms is not well known, there are generic reasons to expect them to have quite varied kinematic structure, as different driving forces should produce winds with correspondingly different acceleration scale lengths and terminal velocities.  Additionally, magnetocentrifugal winds conserve angular velocity along streamlines, as opposed to specific angular momentum as in the case of thermal or radiation driven outflows. Thus, magnetocentrifugal winds would be expected to have higher values of $v_\phi/v_l$ in the line formation region compared to other driving mechanisms (all other parameters being equal). Further modelling is needed here, but this argument clearly demonstrates the link between line profile shape and wind driving mechanism. In fact, the presence of outflow-dominated kinematic components in quasar spectra -- necessitating a line formation region where $v_l > v_\phi$ --  could already be pointing towards, say, line-driven winds as the origin of these emission lines. 

One specific aspect of the line profile shape that changes with varying $\alpha$ (and also $R_v$ and $\dot{M}_w$; see Fig.~\ref{fig:step_par}) is whether or not the line is single-peaked. Our work here demonstrates that disc winds can produce both double- and single-peaked lines, with the observed profile depending on both inclination and the wind parameters. The basic mechanism for the single-peaked behaviour is, once again, the formation of the emission line in an outflow dominated region where, approximately, $v_l > v_\phi$, although the details also depend on the projected velocity and resonant line scattering. We found similar behaviour in our modelling of disc winds in accreting white dwarfs. There the line formation region moved up in the flow as $R_v$ was increased and could cause a transition from double- to single-peaked Balmer lines. This behaviour is distinct from the oft-discussed mechanism proposed by \cite{murray_wind-dominated_1996,murray_disk_1997}, in which radial velocity {\sl gradients} modify the escape probabilities and lead to a single-peaked profile. 
 
The fact that line profile shape depends so clearly on the wind kinematics rather begs the question: what kind of outflows could better mimic the observed broad emission line shapes? We cannot conclusively answer this question at this stage, but it would seem reasonable that alternative prescriptions for quasar outflows could produce a peak of emission that is redward of the blueshift velocity. It would be interesting in future work to consider outflows with more complex kinematics and multiphase structures. One example is the `quasar rain' model suggested by \cite{elvis_quasar_2017}. Alternatively, there are various `failed wind' scenarios which have been explored in both dust-driven \citep{czerny_origin_2011,galliani2013,czerny2017,naddaf2022,naddaf2023} and line-driven \citep{proga_dynamics_2004,giustini_global_2019} flows, which are also relevant to the two-component BLR discussed in the next subsection. These more complex or clumpy wind geometries can represent challenges to the MCRT methods we have used, but are clearly worthy of further investigation.  

\subsection{\civ\ emission space, the BAL connection, and the BLR}
\label{sec:discuss_bals}
In Section~\ref{sec:velocity_bals}, we showed that BALs and blueshifted emission lines could be produced by the same disc wind structure, which is interesting in the context of the results of \citetalias{rankine_bal_2020}. Using an emission-line reconstruction scheme, \citetalias{rankine_bal_2020} found that BALs and non-BALs overlap across much of \civ\ emission space and also demonstrated that the BAL troughs with the highest maximum velocities tended be found at high \civ\ blueshifts. The idea that BALs and emission lines are formed in the same disc wind structure has been explored extensively in various `unification' schemes and modelling efforts \citep[e.g.][]{murray_accretion_1995,elvis_structure_2000,matthews_testing_2016}, but our work explicitly demonstrates that MCRT simulations can simultaneously produce not just emission lines and BALs, but also the blueshifted asymmetry of the emission lines. 
However, there are various caveats and complexities to consider in addition to the skew problem already discussed in Section~\ref{sec:skew}. Inspection of Figs.~\ref{fig:bal} and \ref{fig:bal2} reveals that the line profiles close to BAL viewing angles look quite different to those at low inclinations. Thus, just because BALs and blueshifts are formed in the same outflow does not mean those BAL and non-BAL systems will have similar observed emission line spectra, because of the inclination dependence of, e.g. line widths, EWs and blueshifts. A possible solution here is to consider a more filamentary or spoke-like structure for the wind model, as discussed by \cite{yong_using_2018}. 

In our work, we have focused almost exclusively on the \civline\ emission line. This is partly for simplicity, and partly because the blueshifted asymmetry in this line is the cleanest, strongest and best-studied signature of its type in quasar spectra. We presented full UV spectra for two models in Figs.~\ref{fig:bal} and \ref{fig:bal2}, in which various UV lines such as Lyman-$\alpha$, \nvline, \heiiuv, \sivline\ and \ciiiline\ can be clearly seen, but we have not discussed the kinematic signatures in these lines in detail. Such signatures are astrophysically relevant -- for example, \cite{temple_high-ionization_2021} show that, in quasars with high \civ\ blueshifts, the \ion{Si}{iv}, Lyman-$\alpha$ and \nv\ lines all have blueshifted asymmetric structures resembling \civ. In addition, the \civ\ blueshift is often defined with respect to a line like \mgiiline\, which typically has a centroid close to the systemic velocity. A detailed investigation of the range of emission lines profile shapes in wind models would be an interesting avenue for future work. Nevertheless it does seem fairly clear from the line ratios and profile shapes in Figs.~\ref{fig:bal} and \ref{fig:bal2} that it is quite challenging to get a biconical wind mode of the type considered here to reproduce the full spectrum of a high-blueshift `windy' quasar, even if disc winds can imitate `locally optimally emitting' behaviour \citep{baldwin_locally_1995} by spanning a range of densities and ionization states, as suggested by \citetalias{matthews_stratified_2020}.

It is probably more reasonable to think of the wind-formed emission lines simulated here as corresponding to one component of a multi-faceted broad-line region. If we consider the simplest possible multi-component BLR, the degree of blueshifted asymmetry (and the location in \civ\ emission space) would be partly driven by the relative strength of the `windy' and `peaky' kinematic components \citep[see e.g.][]{temple_high-ionization_2021,temple_testing_2023}. The physical origin of these two components could then correspond to a line- or magnetically driven disc wind, as considered here, and a classic BLR-like component -- perhaps a rotationally dominated turbulent plasma, failed wind or population of clouds -- that produces a symmetric line. This two-component BLR is clearly a toy model; it seems more likely that the BLR is a stratified structure, with complex, probably turbulent, kinematics. Further modelling combined with novel data analysis (for example, attempts to decompose the kinematic components) will thus be essential to reveal its true nature.

\subsection{The opaque midplane}
\label{sec:discuss_midplane}
One of the key ingredients for forming blueshifts in our modelling is the presence of an opaque midplane out to large radii, such that $R_{\rm mid}$ exceeds the characteristic radius of line formation. It was noted by \cite{wilkes1984} nearly four decades ago that the obscuration of the red wing could be due to {\sl ``...the presence of a large disc of gas and dust, probably an extension of gas accreting on to a central black hole''}, extending out to $\sim 10^{18}~{\rm cm}$ ($\sim 1/3~{\rm pc})$. Remarkably, this is the same value of $R_{\rm mid}$ at which blueshifts start to form in the top left panel of Fig.~\ref{fig:step_par}, as dictated by the location of emission line formation. The critical value of $R_{\rm mid}$ likely depends somewhat on the specific model, since we know from Fig.~\ref{fig:lineform} that the line formation region moves around in the wind (in both physical and velocity space) for different wind parameters. Nevertheless, it is interesting to compare this radius to some characteristic radii in the quasar system: for example, taking a $10^{9}~{\rm M_\odot}$ black hole accreting at $\sim 20\%$ of Eddington, the dust sublimation radius is $R_{\rm dust} \sim 1~{\rm pc}$ \citep{barvainis_hot_1987}  and the self-gravitation radius is $R_{\rm sg} \sim 0.02~{\rm pc}$ \citep{King2016}. Thus, $R_{\rm sg}$ is significantly less than, and $R_{\rm dust}$ is broadly comparable to, the required opaque midplane extent. This result is supported observationally by the size of the BLR being similar to $R_{\rm dust}$ in many AGN including high mass quasars 3C 273 and IRAS 13349$+$2438 \citep{Gandhi2015}. We note, however, that it is not clear whether the BLR radius estimated from reverberation mapping is relevant to the line formation radius for the \civ\ blueshifted component in quasars. Indeed, \cite{temple_testing_2023} argue, based on a need for high ionizing fluxes, that the `windy' component of a two-component quasar BLR is formed much closer to the central black hole.

If the line formation radius of the blueshifted \civ\ is comparable to that in this work, it is fairly likely that the obscurer is a self-gravitating, dusty disc or vertically extended torus, rather than a geometrically thin, optically thick accretion disc. In this case, the obscuring midplane may also be playing a role in forming the blueshifts sometimes observed in the narrow lines. In addition, a dusty midplane would have a frequency-dependent opacity. Alternatively, it may be that dust opacity within or surrounding the wind itself has a role to play in absorbing the red wing of broad emission lines, particularly if the wind is dust-opacity driven. Thus, while the presence of an extended obscuring disc is a fairly natural consequence of the axisymmetric quasar system, future work might focus on incorporating dust radiative transfer through a physically motivated model for a dusty disc (and/or dusty outflow), rather than simply treating the midplane as a `brick wall'. 

\subsection{Interpretation of line asymmetries}

We close this section by making some final, rather general comments on the interpretation of asymmetries in emission lines, which are relevant to a host of astrophysical settings. Blueshifted asymmetries in emission lines are not smoking guns for outflows -- they tell us that the line-emitting material is moving towards us, but, unlike BALs, they do not tell us whether that material is located between the observer and the continuum source. Indeed, in our work the disc wind was able to produce both blue- and red-shifted emission line signatures from the same wind geometry, highlighting the potential for confusion. 

Velocity-resolved reverberation mapping results provide important constraints on the kinematics of the BLR \citep[e.g.][]{peterson_reverberation_1993,denney_diverse_2009,pancoast_geometric_2011,waters_reverberation_2016,mangham_reverberation_2017,de_rosa_velocity-resolved_2018}, although most studies focus on local, lower luminosity AGN, which are very different systems to the high-blueshift quasar population. Nevertheless, results can be informative for our work; for example, in a recent study, \cite{oknyansky_multiwavelength_2021} find spectacular asymmetric lines during a changing-look event in the Seyfert galaxy NGC 3516. At first glance, the asymmetries in H$\alpha$ and H$\beta$ they observe have very similar shapes to some of the blueshifts we have found in our models. It is therefore tempting to interpret those asymmetries as coming from a disc wind driven during the changing-look event. However, the reverberation lags in the lines have a red-leads-blue signature which is characteristic of inflow \citep{welsh_echo_1991}. The picture is complicated even further, because recombination lines in particular can have complex responses and transfer functions \citep{korista_what_2004,mangham_reverberation_2017,mangham_reverberation_2019} which can even lead to red-leads-blue signatures from outflows, as shown by \cite[][see also \citealt{yong_kinematics_2017,waters_reverberation_2016}]{mangham_reverberation_2017}. The \cite{oknyansky_multiwavelength_2021}  results are in some sense cautionary, since the NGC 3516 data shows one signature (emission line blueshift) commonly interpreted as evidence for an outflow, and one signature (red-leads-blue reverberation) commonly interpreted as evidence for an inflow.  

Given all the above, one might legitimately ask: are blue-shifted emission lines telling us anything at all about accretion disc winds? While it is true they cannot be considered `smoking guns' of outflowing material in the same way as blue-shifted absorption lines, there are nevertheless at least three pieces of evidence that point towards their formation in outflows. The first is simply physical plausibility; we, and others \citep{richards_broad_2002,chajet_magnetohydrodynamic_2013,chajet_magnetohydrodynamic_2017,yong_kinematics_2017,gravity_collaboration_spatially_2020} have demonstrated that a reasonable disc wind geometry can produce blue-shifted asymmetries. We do not know of an alternative, plausible, physically motivated model for their formation\footnote{We note, however, that \cite{gaskell_line_2013,gaskell_case_2016} argue on mostly empirical grounds that there are a number of problems with a disc wind model for the BLR, instead favouring an inflow model in which blueshifts are created by scattering.} (for example, in a plasma whose kinematics are dominated by rotation or inflow). The second piece of evidence is the behaviour of BAL properties in \civ\ emission space. \citetalias{rankine_bal_2020} find that the most extreme BALs (i.e. highest Balnicity index [BI] and largest maximum trough velocities) are typically associated with the largest \civ\ blueshifts when the spectrum is reconstructed using mean-field independent component analysis. While this relationship does not necessitate a direct physical association, the simplest explanation is that \civ\ blueshifts are created by the BAL outflows themselves. Finally, there is the general point that, from the prevalence of \civ\ BALs, we know quasars produce outflows carrying  a significant \civ\ ion population, so it seems rather natural to associate these outflows with the \civ\ emission lines. This argument is strengthened by our work, since we have shown that \civ\ BALs and emission line blueshifts can be formed in the same disc wind under reasonable assumptions (although see also discussion in Section~\ref{sec:discuss_bals}).

\subsection{Limitations and model assumptions}
\label{sec:limitations}

Our work has a number of limitations, which we briefly discuss as well as referring the reader to our earlier work (\citealt{matthews_testing_2016};
\citetalias{matthews_stratified_2020}). 

Given the obvious impact of wind geometry and the velocity law on the line profile shape, it seems clear that adopting the \cite{shlosman_winds_1993} wind parametrization is a limitation of this work. In particular, it may be possible to get a better match to observed quasar spectra with a different wind model. It would be  valuable to test model prescriptions which are more physically motivated, in the sense that the wind streamlines and opening angles are obtained from a self-consistent solution of the wind dynamics. Similarly, although our choices of variable wind parameters and the values of the grid points in our simulation grid are sensible, there are also feasibly other model parameters or regions of parameter space that should be investigated.

In terms of the physics in the simulations, the main limitations are probably the treatment of clumping and the SED. Our `microclumping' treatment, in which inhomogeneities in the flow are implemented through a single volume filling factor, is necessarily simplistic and clearly does not capture the physics of multiphase, clumpy outflows. In terms of the SED, we have assumed an isotropic SED, and while we have investigated the sensivitity of the \civ\ line profile to both the SED shape and anisotropy (Section~\ref{sec:sed_impact}), the treatment of the continuum radiation field can dramatically affect the ionization state of the wind and the EW of the lines in the synthetic spectra. 

\section{Conclusions}
\label{sec:conclusions}
We have demonstrated how blueshifts can form in biconical disc winds (e.g. Figs.~\ref{fig:heuristic} and \ref{fig:blueshift_grid}), and used Monte Carlo radiative transfer and photoionization simulations to examine trends with inclination, wind geometry, and other wind parameters. Beyond this result, our main conclusions are as follows:
\begin{enumerate}
    \item We find that a key condition for blueshift formation is that the disc mid-plane is opaque out to radii beyond the line formation region, so that the receding wind bicone is obscured (section~\ref{sec:heuristic}; Figs.~\ref{fig:heuristic}, \ref{fig:blueshift_grid} and \ref{fig:step_par}). 
    \item Blueshifts are only produced at relatively low (that is, more face on) inclinations, with $\approx70$\,per cent of the spectra with \civ\ blueshifts seen at $\theta_i \lesssim 20^\circ$, and $\approx95$\,per cent at $\theta_i \lesssim 40^\circ$ (section~\ref{sec:inclination}; Fig.~\ref{fig:inclination}). 
    \item We find that a combination of ionization and emission measure effects act to confine the \civ\ line formation region to near the base of the outflow (section~\ref{sec:velocity_bals}; Fig.~\ref{fig:phase}), where velocities are typically $300-3000~{\rm km~s}^{-1}$. Further out in the flow, densities are too low, and the \civline\ emissivity drops dramatically. This mechanism provides a natural explanation for the size of the observed \civ\ blueshifts and shows that emission line velocities do not necessarily provide any information about the terminal velocity in the flow. It also suggests that higher \civ\ blueshifts would require higher mass-loss rates or lower volume filling factors, such that the lines could form at higher velocity.
    \item We demonstrate that BALs and emission line blueshifts can be formed in the same outflow (section~\ref{sec:velocity_bals}; Figs.~\ref{fig:bal} and ~\ref{fig:bal2}), with BALs observed at higher inclinations when the viewing angle intersects a significant portion of the wind cone $(\theta_i \gtrsim \theta_{\rm min})$, and emission line blueshifts at low inclinations $(\theta_i \lesssim \theta_{\rm min})$. This result has interesting implications for our understanding of the BAL versus non-BAL dichotomy and, in particular, the \civ\ emission space. However, it is challenging to construct a workable biconical model in which the emission line properties in the BAL and non-BAL quasars are as similar as in the observational data (\citetalias{rankine_bal_2020}).
    \item Despite our success in reproducing blueshifted emission lines within a disc wind model, we cannot match the detailed line profile shapes (section~\ref{sec:skew}). In particular, the model spectra tend to be `positively skewed' in the sense that the line peak lies blueward of the blueshift calculated from the wavelength bisecting the line flux; observed spectra display the opposite behaviour. This discrepancy may suggest a fundamental problem with a disc wind origin for \civ\ blueshifts, but it is probably more likely that our biconical model is limited in its parametrization and a poor imitation of a real, dynamic quasar wind.  
    \item Redshifted emission lines can also be produced when the wind region on the far side of the black hole is viewed at such an angle that the plasma is receding from the observer (section~\ref{sec:inclination}). This starts to occur above a critical viewing angle of $\theta_{\rm red} \approx 90^\circ - \theta_{\rm min}$, a necessary but not sufficient criterion for redshifted line emission. Radiative transfer effects (i.e. preferential absorption of the blue wind) within the wind are critical for forming these asymmetries even when $\theta_i \gtrsim \theta_{\rm red}$.
    \item We find that the shape of the velocity law has a significant impact on the shape of the \civ\ line profile, which we demonstrate through variation of the acceleration exponent $\alpha$ (section~\ref{sec:wind_params}; Figs.~\ref{fig:step_par} and \ref{fig:lineform}). The velocity law has a two-fold effect on the line profile by changing the density structure, and therefore location of the line forming region in velocity space, and also altering the projected velocity into the line of sight. A critical factor is how quickly $v_l$ increases along streamlines with respect to the decrease of $v_\phi$. Importantly, the impact of the wind velocity profile suggests that emission line blueshifts and asymmetries can be used as a probe of disc wind driving mechanisms in quasars and other accreting systems. 
\end{enumerate}
Future work should move beyond the limited parameterisation of a disc wind we adopted, and, in an ideal world, construct {\sl ab initio} models with different driving physics. When confronted with observations, such an approach will help establish how blueshifted emission lines are formed and more generally help understand the physical mechanisms driving the diversity of line profile shapes and associated BLR kinematics. Regardless of the driving mechanism, accurate radiative transfer calculations will be critical in order to produce consistent synthetic spectra. In addition, our work suggests that the emission lines can in principle be a powerful probe of the underlying physics of outflows/winds in quasars and other accreting systems.

\section*{Data Availability}
The bulk of the simulation data used in this article is public in a github repository at \url{https://github.com/jhmatthews/blueshifts},  with an associated Zenodo record and digital object identifier \citep{zenodo}. The observational data underlying this article come from the Sloan Digital Sky Survey Data Release 16. The composite quasar spectra are available in the Supporting Information included with \citet{stepney}. Any other derived data products are available from the authors upon reasonable request. The \python\ radiative transfer code is an open-source project available at \url{https://github.com/agnwinds/python}. 

\section*{Acknowledgements}
We thank the anonymous referee for reading the manuscript and for their constructive comments. We woulda also like to thank Andrew Sellek, Stuart Sim, Nico Scepi and Teo Mu\~{n}oz-Darias for helpful discussions. JHM acknowledges support from a Royal Society University Research Fellowship and, previously, the Herchel Smith fund. ALR acknowledges support from a UKRI Future Leaders Fellowship (grant code: MR/T020989/1). MJT acknowledges support from a FONDECYT postdoctoral fellowship (3220516). This work was performed using resources provided by the Cambridge Service for Data Driven Discovery (CSD3) operated by the University of Cambridge Research Computing Service (www.csd3.cam.ac.uk), provided by Dell EMC and Intel using Tier-2 funding from the Engineering and Physical Sciences Research Council (capital grant EP/T022159/1), and DiRAC funding from the Science and Technology Facilities Council (www.dirac.ac.uk). We gratefully acknowledge the use of the following software packages: OpenMPI \citep{openmpi}, matplotlib \citep{matplotlib}, astropy \citep{astropy_collaboration_astropy_2013,astropy_collaboration_astropy_2018}. For the purpose of open access, the authors have applied a Creative Commons Attribution (CC BY) licence to any Author Accepted Manuscript version arising from this submission.

Funding for the Sloan Digital Sky 
Survey IV has been provided by the 
Alfred P. Sloan Foundation, the U.S. 
Department of Energy Office of 
Science, and the Participating 
Institutions. 
SDSS-IV acknowledges support and 
resources from the Center for High 
Performance Computing  at the 
University of Utah. The SDSS 
website is www.sdss4.org.
SDSS-IV is managed by the 
Astrophysical Research Consortium 
for the Participating Institutions 
of the SDSS Collaboration including 
the Brazilian Participation Group, 
the Carnegie Institution for Science, 
Carnegie Mellon University, Center for 
Astrophysics | Harvard \& 
Smithsonian, the Chilean Participation 
Group, the French Participation Group, 
Instituto de Astrof\'isica de 
Canarias, The Johns Hopkins 
University, Kavli Institute for the 
Physics and Mathematics of the 
Universe (IPMU) / University of 
Tokyo, the Korean Participation Group, 
Lawrence Berkeley National Laboratory, 
Leibniz Institut f\"ur Astrophysik 
Potsdam (AIP),  Max-Planck-Institut 
f\"ur Astronomie (MPIA Heidelberg), 
Max-Planck-Institut f\"ur 
Astrophysik (MPA Garching), 
Max-Planck-Institut f\"ur 
Extraterrestrische Physik (MPE), 
National Astronomical Observatories of 
China, New Mexico State University, 
New York University, University of 
Notre Dame, Observat\'ario 
Nacional / MCTI, The Ohio State 
University, Pennsylvania State 
University, Shanghai 
Astronomical Observatory, United 
Kingdom Participation Group, 
Universidad Nacional Aut\'onoma 
de M\'exico, University of Arizona, 
University of Colorado Boulder, 
University of Oxford, University of 
Portsmouth, University of Utah, 
University of Virginia, University 
of Washington, University of 
Wisconsin, Vanderbilt University, 
and Yale University.


\appendix
\section{Model parameters and Additional Plots}
\label{appendixa}
\begin{figure*}
    \centering
    \includegraphics[width=0.49\linewidth]{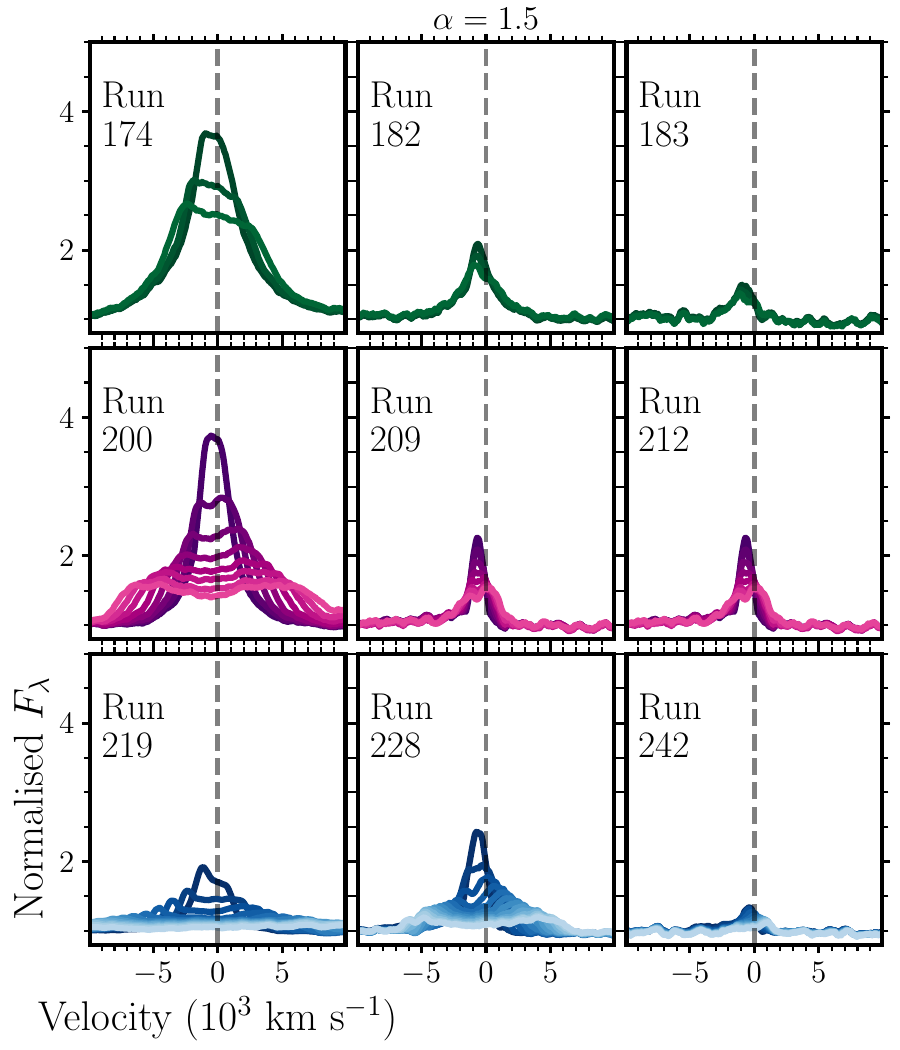}
    \includegraphics[width=0.49\linewidth]{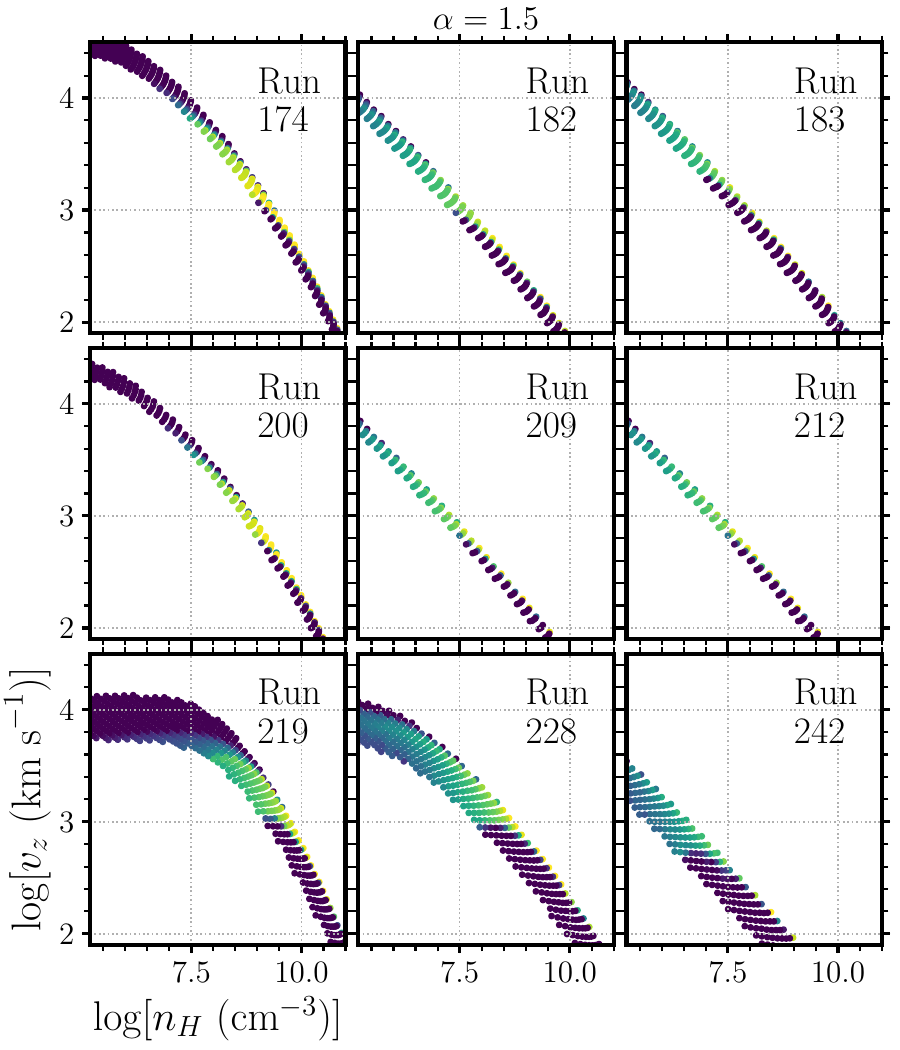}
    \caption{{\sl Left:} As Fig.~\ref{fig:blueshift_grid} but for only the nine corresponding $\alpha=1.5$ models (full parameters for each model given in Table~\ref{tab:params}). The colour scheme is the same as in Fig.~\ref{fig:blueshift_grid} and denotes wind opening angle or geometry, with the colour map itself encoding inclination. {\sl Right:} As Fig.~\ref{fig:phase} but for only the nine corresponding $\alpha=1.5$ models. The colour map matches Fig.~\ref{fig:phase} and shows \civline\ line luminosity.}
    \label{fig:alpha1.5_phase}
\end{figure*}

\begin{figure*}
    \centering
    \includegraphics[width=0.49\linewidth]{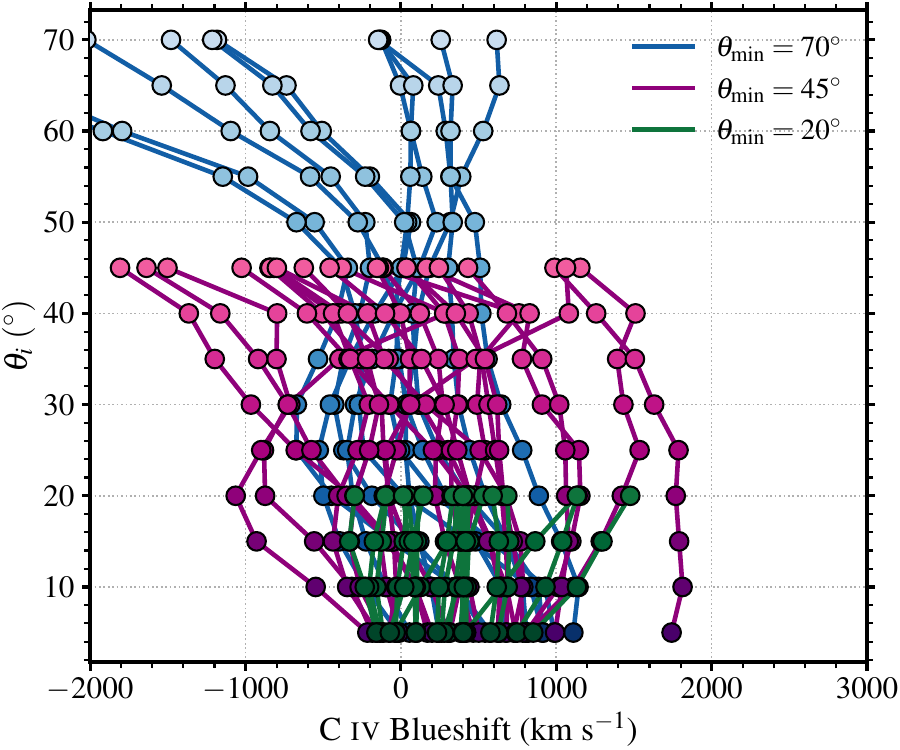}
    \includegraphics[width=0.49\linewidth]{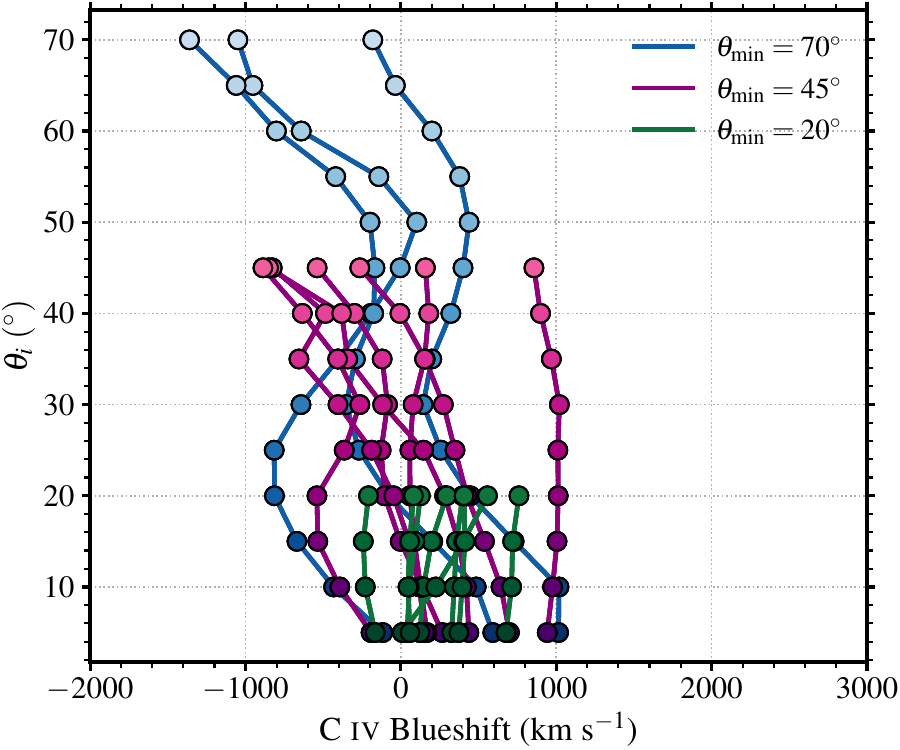}
    \caption{Blueshift as a function of inclination for the three considered wind geometries (as in the left-hand panel of Fig.~\ref{fig:inclination}), for $\alpha=1$ (left) and $\alpha=1.5$ (right).}
    \label{fig:inclination_appendix}
\end{figure*}

In Table~\ref{tab:params}, we present the model parameters for the 18 members of the simulation grid shown in Figs.~\ref{fig:blueshift_grid} and ~\ref{fig:phase}, together with labels matching those in the figure panels. 
In addition we show the nine models with $\alpha=1.5$ with the other parameters matching the corresponding  $\alpha=0.5$  and $\alpha=1$ models. 
Fig.~\ref{fig:alpha1.5_phase} shows the line profiles (left panel) and wind structure in $n_H-v_z$ space (right panel) from these nine models with $\alpha=1.5$. Finally, in Fig.~\ref{fig:inclination_appendix}, we show \civ\ blueshift as a function of inclination for the $\alpha=1$ and $\alpha=1.5$ models, analogously to the left-hand panel of Fig.~\ref{fig:inclination}.

\begin{table}
\centering
\begin{tabular}{ccccccc}
\hline 
Run Number& $\theta_{\rm min}$ & $\dot{M}_w~(M_{\odot}~{\rm yr}^{-1})$ & $R_v$~(cm) & $f_\infty$ & $\alpha$ & Label \\ 
 \hline 
Run 12 & $20$ & $10$ & $10^{ 18 }$ & $1$ & $0.5$ & a) \\ 
Run 20 & $20$ & $5$ & $10^{ 19 }$ & $1$ & $0.5$ & b) \\ 
Run 21 & $20$ & $10$ & $10^{ 19 }$ & $1$ & $0.5$ & c) \\ 
Run 38 & $45$ & $5$ & $10^{ 18 }$ & $1$ & $0.5$ & d) \\ 
Run 47 & $45$ & $5$ & $10^{ 19 }$ & $1$ & $0.5$ & e) \\ 
Run 50 & $45$ & $5$ & $10^{ 19 }$ & $2$ & $0.5$ & f) \\ 
Run 57 & $70$ & $10$ & $10^{ 17 }$ & $1$ & $0.5$ & g) \\ 
Run 66 & $70$ & $10$ & $10^{ 18 }$ & $1$ & $0.5$ & h) \\ 
Run 80 & $70$ & $5$ & $10^{ 19 }$ & $3$ & $0.5$ & i) \\ 
Run 93 & $20$ & $10$ & $10^{ 18 }$ & $1$ & $1$ & j) \\ 
Run 101 & $20$ & $5$ & $10^{ 19 }$ & $1$ & $1$ & k) \\ 
Run 102 & $20$ & $10$ & $10^{ 19 }$ & $1$ & $1$ & l) \\ 
Run 119 & $45$ & $5$ & $10^{ 18 }$ & $1$ & $1$ & m) \\ 
Run 128 & $45$ & $5$ & $10^{ 19 }$ & $1$ & $1$ & n) \\ 
Run 131 & $45$ & $5$ & $10^{ 19 }$ & $2$ & $1$ & o) \\ 
Run 138 & $70$ & $10$ & $10^{ 17 }$ & $1$ & $1$ & p) \\ 
Run 147 & $70$ & $10$ & $10^{ 18 }$ & $1$ & $1$ & q) \\ 
Run 161 & $70$ & $5$ & $10^{ 19 }$ & $3$ & $1$ & q) \\ 
\hline
Run 174 & $20$ & $10$ & $10^{ 18 }$ & $1$ & $1.5$ &  \\ 
Run 182 & $20$ & $5$ & $10^{ 19 }$ & $1$ & $1.5$ &  \\ 
Run 183 & $20$ & $10$ & $10^{ 19 }$ & $1$ & $1.5$ &  \\ 
Run 200 & $45$ & $5$ & $10^{ 18 }$ & $1$ & $1.5$ &  \\ 
Run 209 & $45$ & $5$ & $10^{ 19 }$ & $1$ & $1.5$ &  \\ 
Run 212 & $45$ & $5$ & $10^{ 19 }$ & $2$ & $1.5$ &  \\ 
Run 219 & $70$ & $10$ & $10^{ 17 }$ & $1$ & $1.5$ &  \\ 
Run 228 & $70$ & $10$ & $10^{ 18 }$ & $1$ & $1.5$ &  \\ 
Run 242 & $70$ & $5$ & $10^{ 19 }$ & $3$ & $1.5$ &  \\ 

\hline
\end{tabular}
\caption{
{\sl Top panel:} Model parameters for the 18 models presented in Figs.~\ref{fig:blueshift_grid} and \ref{fig:phase}, with labels a) to q) matching those in the figure panels. {\sl Bottom panel:} corresponding $\alpha=1.5$ models shown only in the appendix (Fig.~\ref{fig:alpha1.5_phase}).
}
\label{tab:params}
\end{table}

\bsp	
\label{lastpage}
\end{document}